\documentclass[nofootinbib,11pt]{revtex4-1}
\usepackage{amsmath,amssymb,pgf,pgfarrows,pgfnodes,float,appendix, hyperref,slashed,breakurl,graphicx}
\definecolor{Color}{rgb}{0.28, 0.24, 0.55}
\definecolor{Orange}{rgb}{1,0.38,0.11}
\hypersetup{
    colorlinks = true,
    citecolor = Orange,
    linkcolor = Color,
    urlcolor  = Orange,
}
\usepackage{graphicx}
\usepackage{subfigure}
\usepackage[margin=0.9in]{geometry}

\usepackage{enumerate}

\newcommand{\SU}{\text{SU}}

\usepackage{color, colortbl}
\definecolor{Gray}{gray}{0.8}
\definecolor{GrayLight}{gray}{0.4}
\definecolor{Darkgreen}{RGB}{30,120,30}
\definecolor{granate}{rgb}{0.8039,0.2,0.2}
\newcommand{\beq}{\begin{equation}}
\newcommand{\eeq}{\end{equation}}
\newcommand{\bea}{\begin{eqnarray}}
\newcommand{\eea}{\end{eqnarray}}

\usepackage{tikz}
\usetikzlibrary{"arrows", "automata", "backgrounds", "calendar", "chains", "matrix", "mindmap", "patterns", "petri", "shadows", "shapes.geometric", "shapes.misc", "spy", "trees"}
\usetikzlibrary{arrows,shapes}
\usetikzlibrary{trees}
\usetikzlibrary{matrix,arrows} 				
\usetikzlibrary{positioning}				
\usetikzlibrary{calc,through}				
\usetikzlibrary{decorations.pathreplacing}  
\usepackage{pgffor}							

\usetikzlibrary{decorations.pathmorphing}	
\usetikzlibrary{decorations.markings}
\tikzset{
    vector/.style={decorate, decoration={snake}, draw},
	provector/.style={decorate, decoration={snake,amplitude=2.5pt}, draw},
	antivector/.style={decorate, decoration={snake,amplitude=-2.5pt}, draw},
    fermion/.style={draw=black, postaction={decorate},
        decoration={markings,mark=at position .55 with {\arrow[draw=black]{>}}}},
    fermioncyan/.style={draw=black, postaction={decorate},
        decoration={markings,mark=at position .55 with {\arrow[draw=cyan]{<}}}},
    fermiondif/.style={draw=black, postaction={decorate},
        decoration={markings,mark=at position .7 with {\arrow[draw=black]{>}}}},
            fermiondif2/.style={draw=black, postaction={decorate},
        decoration={markings,mark=at position .7 with {\arrow[draw=black]{<}}}},
    fermionend/.style={draw=black, postaction={decorate},
        decoration={markings,mark=at position 1 with {\arrow[draw=black]{>}}}},
    fermionuchannel2/.style={draw=black, postaction={decorate},
        decoration={markings,mark=at position .4 with {\arrow[draw=black]{>}}}},
    scalardif/.style={dashed,draw=black, postaction={decorate},
        decoration={markings,mark=at position .7 with {\arrow[draw=black]{>}}}},
    scalarend/.style={dashed,draw=black, postaction={decorate},
        decoration={markings,mark=at position 1 with {\arrow[draw=black]{>}}}},
    fermionbar/.style={draw=black, postaction={decorate},
        decoration={markings,mark=at position .55 with {\arrow[draw=black]{<}}}},
    fermionnoarrow/.style={draw=black},
    gluon/.style={decorate, draw=black,
        decoration={coil,amplitude=4pt, segment length=5pt}},
    scalar/.style={dashed,draw=black, postaction={decorate},
        decoration={markings,mark=at position .55 with {\arrow[draw=black]{>}}}},
    scalarcyan/.style={dashed,draw=black, postaction={decorate},
        decoration={markings,mark=at position .55 with {\arrow[draw=cyan]{>}}}},
    scalaruchannel1/.style={dashed,draw=black, postaction={decorate},
        decoration={markings,mark=at position .7 with {\arrow[draw=black]{>}}}},
                  scalaruchannel2/.style={dashed,draw=black, postaction={decorate},
        decoration={markings,mark=at position .4 with {\arrow[draw=black]{>}}}},
    scalarbar/.style={dashed,draw=black, postaction={decorate},
        decoration={markings,mark=at position .55 with {\arrow[draw=black]{<}}}},
    scalarnoarrow/.style={dashed,draw=black},
    electron/.style={draw=black, postaction={decorate},
        decoration={markings,mark=at position .55 with {\arrow[draw=black]{>}}}},
	bigvector/.style={decorate, decoration={snake,amplitude=4pt}, draw},
}

\tikzstyle{block} = [draw, rectangle, 
    minimum height=3em, minimum width=6em]

\usepackage{tikz}
\usetikzlibrary{fit}
\tikzset{%
  highlight/.style={rectangle,rounded corners,color=granate,draw,text opacity =1,
    fill opacity=0.5,thick,inner sep=0pt}
}

\usepackage{placeins}

\tikzset{
    cross/.pic = {
    \draw[rotate = 45] (-#1,0) -- (#1,0);
    \draw[rotate = 45] (0,-#1) -- (0, #1);
    }
}

\tikzset{
    square/.style={%
        draw=none,
        circle,
        append after command={%
            \pgfextra \draw[#1] (\tikzlastnode.north-|\tikzlastnode.west) rectangle 
                (\tikzlastnode.south-|\tikzlastnode.east);\endpgfextra}
    },
    square/.default=black
}

\tikzstyle{block} = [draw, rectangle, 
    minimum height=3em, minimum width=6em]

\usepackage{xparse}
\NewDocumentCommand\semiloop{O{black}mmmO{}O{above}}
{%
\draw[#1] let \p1 = ($(#3)-(#2)$) in (#3) arc (#4:({#4+180}):({0.5*veclen(\x1,\y1)})node[midway, #6] {#5};)
}

\begin{document}

\title{\Large{Flavor Anomalies and Quark-Lepton Unification}
}
\author{Pavel Fileviez P\'erez$^{1}$, Clara Murgui$^{2}$}
\affiliation{$^{1}$Physics Department and Center for Education and Research in Cosmology and Astrophysics (CERCA), 
Case Western Reserve University, Cleveland, OH 44106, USA \\
$^{2}$Walter Burke Institute for Theoretical Physics, California Institute of Technology, Pasadena, CA 91125}
\email{pxf112@case.edu, cmurgui@caltech.edu}
\vspace{1.5cm}
\begin{abstract}
We show that one can explain the neutral and charged anomalies in $B$-meson decays in the minimal theory for quark-lepton unification.
The implications for flavor violating processes are discussed in detail. Strikingly, experimental observations suggest that the unification of quarks and leptons could be realized at the ${\cal O}(10^2)$ TeV scale.
\end{abstract}

\maketitle 

\section{Introduction}
The Standard Model of Particle Physics (SM), despite the glory of explaining the vast majority of phenomena seen at experiment, cannot be the final theory. There are theoretical, phenomenological and cosmological motivations for a high-energy completion, which leads to reconsider the SM as part of the renormalizable terms of an effective field theory (EFT). Amongst the experimental corroboration for physics beyond the SM, it is well-known that at least two neutrinos must have a non-zero mass and we do not have an explanation for the hierarchy between the charged fermion masses. Besides, an already long-standing tension has been reported in the semileptonic decays of the $B$ meson~\cite{Aaij:2014ora,Aaij:2017vbb,Aaij:2019wad,Aaij:2021vac,Lees:2012xj,Lees:2013uzd,Aaij:2015yra,Huschle:2015rga,Hirose:2016wfn,Hirose:2017dxl,Aaij:2017uff,Aaij:2017deq,Abdesselam:2019dgh}. These discrepancies with respect to the SM prediction on lepton flavor universality, referred in the literature as flavor anomalies, exhibit a hierarchical flavor structure which seems to be correlated with the flavor puzzle in the SM and suggests the presence of new physics at the few TeV scale.

In this paper we discuss the possibility to explain the flavor anomalies in the context of the simplest theory where quarks and leptons are unified at the low scale, following Pati and Salam's idea~\cite{Pati:1974yy}. The theory, presented in Ref.~\cite{FileviezPerez:2013zmv}, is based on $\SU(4)_C \otimes \SU(2)_L \otimes \text{U}(1)_R$ and constitutes one of the simplest extensions of the SM, being only separated by one breaking step from the SM gauge group, containing the SM fermions, the right-handed neutrinos, and being able to describe physics at the multi-TeV scale with light neutrino masses generated by the inverse seesaw mechanism.

The minimal theory for quark-lepton unification~\cite{FileviezPerez:2013zmv} predicts a vector leptoquark, $X_\mu \sim (\mathbf{3},\mathbf{1},2/3)_\text{SM}$, whose mass defines the scale of quark-lepton unification. This scale does not necessary have to be large as their GUT partners $\SU(5)$ or $\text{SO}(10)$ since baryon number is preserved at the renormalizable level and neutrinos get mass through the inverse seesaw mechanism without the need of a GUT scale to suppress the active neutrino masses. It is well known, however, that the vector leptoquark can mediate the decay $K_L \to e^\pm  \mu^\mp$, which sets a lower bound on the quark-lepton unification scale larger than $10^3$ TeV if the mixings between quarks and leptons are neglected~\cite{Valencia:1994cj,Smirnov:2007hv}. Amongst the new field content, the theory predicts four new scalars, which have unique properties: (a) a scalar diquark, 
(b) a second Higgs doublet, responsible to break the degeneracy between the masses for charged leptons and down quarks (the latter being a consequence of quark-lepton unification), (c) and two scalar leptoquarks, $\Phi_3 \sim (\mathbf{\bar 3}, \mathbf{2}, 1/6)_\text{SM}$ and $\Phi_4 \sim (\mathbf{3},\mathbf{2},7/6)_\text{SM}$, which will induce fermion flavor violation through the Yukawa interactions. 

The leptoquarks $\Phi_3$ and $\Phi_4$ are the only pure scalar leptoquarks (not admitting diquark couplings) to which the SM fermions can interact. Their interactions conserve baryon number at the renormalizable level. They are also protected against unacceptable baryon number violation at the non-renormalizable level by the Pati-Salam $\SU(4)$ symmetry and its minimal matter content, forbidding baryon number violating operators until dimension seven in the effective field theory~\cite{Murgui:2021bdy}, which makes them excellent candidates to live at the few TeV scale. The leptoquark interactions are defined by new mixing matrices between the quarks and leptons, which are in principle unknown. However, the theory predicts unique relations between the decay channels of the vector leptoquark, the scalar leptoquarks and the new Higgs boson~\cite{Perez:2021mgz}, which are independent of them once the summation over final fermions is performed, being these very attractive signatures of the quark-lepton unification symmetry. We refer the reader to Refs.~\cite{Popov:2005wz,Biggio:2016wyy,Faber:2018afz,Faber:2018qon,Martynov:2020cjd} for phenomenological studies in the context of this theory.

In this paper, we focus on the flavor violation constraints in the theory. We show how the scalar leptoquark $\Phi_4\sim (\mathbf{3},\mathbf{2},7/6)_\text{SM}$ can accommodate the observed anomalies while being consistent with any other existing experimental constraint. This possibility was already considered in Ref.~\cite{Popov:2019tyc} from a simplified model perspective, with generic Yukawa interactions of this leptoquark with matter, and it was further addressed in Ref.~\cite{Perez:2021ddi} in the context of quark-lepton unification, where the flavor anomalies were also explained by assuming empirically some textures for the physical couplings of leptoquarks with the SM fermions. The main focus in the latter was to simultaneously address the neutral flavor anomalies and the recent (and still ambiguous) anomalous magnetic moment of the muon~\cite{Muong-2:2006rrc,Muong-2:2021ojo} consistently with the existing experimental bounds. Unfortunately, for the $\Phi_3$ contribution to the flavor anomalies studied there, large Wilson coefficients are needed to address them~\cite{Perez:2021ddi}, which lead to large branching ratios for the meson decays that might be in tension with the experiment.  

In this new study we exploit the fact that the Yukawa interactions relevant for the flavor anomalies are strikingly factorized in unitary components and diagonal matrices composed of the physical masses of the SM fermions at the quark-lepton unification scale. The latter allows us to determine the textures given the available experimental constraints and make predictions and correlate observables that allow the testability of the theory in a foreseeable future. We will show how the role of $\Phi_4$ in the neutral anomalies presented in Ref.~\cite{Perez:2021ddi} is recovered from this new point of view of fermion mass spectrum and unitarity and will explore its implications on the rest of flavour interactions. Remarkably, the theory predicts that the contribution of the vector leptoquark to $K_L \to e^\pm  \mu^\mp$ is suppressed, thereby allowing the theory to be realized at a lower scale than that expected in generic Pati-Salam theories, particularly around $100$ TeV. This study suggests a strong correlation between the ratios ${\cal R}_{K^{(*)}}$ and $\text{Br}(\tau \to e \gamma)$ that will allow to test the possibility to address the flavour anomalies in the near future.

Altogether we believe that the minimal theory based on the gauge group $\SU(4)_C \otimes \SU(2)_L \otimes \text{U}(1)_R$ is a very simple and compelling extension of the SM that can address the observed flavor anomalies, unifies quarks and leptons, allows for a GUT embedding at the high scale, and predicts masses for the neutrinos. We insist that there is hope to test its existence in the near future, since non-suppressed Yukawa interactions are needed to correct the fermion mass relations and the scalar leptoquark $\Phi_4$ in this theory cannot be heavier than a few TeVs. Therefore, we eagerly encourage our experimental colleagues to look for the imprints of quark-lepton unification discussed in this paper in the flavor and collider experiments.

\section{Minimal Theory for Quark-Lepton Unification}
\label{sec:Theory}
A simple renomalizable theory for quark-lepton unification at the low scale was proposed in Ref.~\cite{FileviezPerez:2013zmv}, which can be seen as a low energy limit of the Pati-Salam theory.
This theory is based on the gauge symmetry: $$\SU(4)_C \otimes \SU(2)_L \otimes \text{U}(1)_R,$$ and each family of the SM matter fields plus three right-handed neutrinos are unified in three representations:
\begin{equation}
F_{QL} =\begin{pmatrix} u & & \nu \\ d & & e \end{pmatrix}  
 \sim (\mathbf{4}, \mathbf{2}, 0), 
\quad
F_u = \begin{pmatrix} u^c & \nu^c \end{pmatrix}
 \sim (\mathbf{\bar{4}}, \mathbf{1}, -1/2), 
\quad {\rm{and}} \quad
F_d = \begin{pmatrix} d^c  & e^c \end{pmatrix}
 \sim (\mathbf{\bar{4}}, \mathbf{1}, 1/2), \nonumber
\end{equation}
while the $\SU(4)_C$ gauge fields live in $A_\mu \sim (\mathbf{15}, \mathbf{1},0)$. The minimal Higgs sector is composed of three scalar representations: $\Phi \sim (\mathbf{15}, \mathbf{2}, 1/2)$, $\chi \sim  (\mathbf{4}, \mathbf{1}, 1/2)$ and $H_1 \sim  (\mathbf{1}, \mathbf{2}, 1/2)$. 
The Yukawa interactions for the charged fermions can be written as
\begin{eqnarray}
- {\cal L} &\supset&
Y_1  \,  {F}_{QL}^a \epsilon_{ab} H_1^b  F_u   \ + \ Y_2 \, {F}_{QL}^a \epsilon_{ab} \Phi^b  F_u 
+   Y_3 \,  H_{1}^\dagger  {F}_{QL} F_d  \ + \  Y_4  \, \Phi^\dagger {F}_{QL}  F_d   + \mbox{h.c.},
\label{eq:Yukawa}
\end{eqnarray}
where $a$ and $b$ correspond to the $\SU(2)_L$ indices, while for the neutrinos one can implement the inverse seesaw mechanism using the terms
\begin{eqnarray}
- {\cal L} &\supset&
Y_5 F_u \chi S  \ + \  \frac{1}{2} \mu S S   + \mbox{h.c.}.
\end{eqnarray} 
Here the fields $S \sim (\mathbf{1}, \mathbf{1}, 0)$ are SM fermionic singlets.

This theory predicts a vector leptoquark, $X_\mu \sim (\mathbf{3},\mathbf{1},2/3)_\text{SM}$,\footnote{We use the label ``SM" when referring to the quantum numbers of the SM gauge group, i.e. $\SU(3)_C \otimes \SU(2)_L \otimes \text{U}(1)_Y$. Otherwise, the quantum charges will refer to the Pati-Salam gauge symmetry $\SU(4)_C \otimes \SU(2)_L \otimes \text{U}(1)_R$.} associated to the $\SU(4)_C$ symmetry breaking, a scalar diquark $\Phi_8 \sim (\mathbf{8},\mathbf{2},1/2)_\text{SM}$, a second Higgs doublet $H_2 \sim (\mathbf{1},\mathbf{2},1/2)_\text{SM}$, and two physical scalar leptoquarks $\Phi_3 \sim (\mathbf{\bar{3}}, \mathbf{2},-1/6)_\text{SM}$ and $\Phi_4 \sim (\mathbf{3}, \mathbf{2}, 7/6)_{\rm SM}$. All these scalars live in the adjoint representation $\Phi \sim (\mathbf{15}, \mathbf{2}, 1/2)$ and therefore interact with the SM fermions through the Yukawa interaction in Eq.~\eqref{eq:Yukawa}. The scalar leptoquarks  decompose in $\SU(2)_L$ components as,
\begin{equation}
\Phi_3 =  \begin{pmatrix} \phi_3^{1/3} \\[1ex]\phi_3^{-2/3} \end{pmatrix},
 \hspace{0.5cm} \text{and} \hspace{0.5cm} \Phi_4 =\begin{pmatrix} \phi_4^{5/3} \\[1ex]\phi_4^{2/3} \end{pmatrix},
\end{equation}
where the numbers in the superscript denote the electric charge. The Yukawa interactions for $\Phi_3$ and $\Phi_4$ are obtained by expanding the interactions in Eq.~\eqref{eq:Yukawa} in terms of the $\SU(3)_C \otimes \SU(2)_L \otimes \text{U}(1)_Y$ fields,
\begin{eqnarray}
-\mathcal{L} &\supset& Y_2 \, \varepsilon_{ab} \, \ell_L^a \, \Phi_4^b \, (u^c)_L + Y_2 \, \varepsilon_{ab} \, Q_L^a \, \Phi_3^b \, (\nu^c)_L 
+  Y_4 \, \Phi_3^\dagger \, \ell_L \, (d^c)_L + Y_4  \, \Phi_4^\dagger  \, Q_L \, (e^c)_L     + \mbox{h.c.} \, .
\end{eqnarray}
Notice that in this sector the interactions are parametrized by only two Yukawa matrices because the $\SU(4)_C$ symmetry relates the different Yukawa interactions in a unique way, as Eq.~\eqref{eq:Yukawa} displays.
After spontaneous breaking of the electroweak symmetry, the charged fermion masses and the Dirac neutrino masses ($M_\nu^D$) are given by
\begin{equation}
\begin{split}
 M_U &= \displaystyle Y_1 \frac{ v_1}{\sqrt{2}} + \frac{1}{2 \sqrt{3}} Y_2 \frac{ v_2}{\sqrt{2}}, \qquad  M_\nu^D = Y_1 \frac{ v_1}{\sqrt{2}} - \frac{\sqrt{3}}{2} Y_2 \frac{ v_2}{\sqrt{2}}, \\
M_D &= \displaystyle Y_3  \frac{v_1}{\sqrt{2}} + \frac{1}{2 \sqrt{3}} Y_4 \frac{ v_2}{\sqrt{2}}, \qquad  \,  M_E = Y_3\frac{ v_1}{\sqrt{2}} - \frac{\sqrt{3}}{2} Y_4 \frac{ v_2}{\sqrt{2}}.
\end{split}
\label{eq:massesYuk}
\end{equation}
Here the VEVs of the Higgs doublets are defined as $\langle H^0_1 \rangle = v_1 / \sqrt{2}$, and $\langle H^0_2  \rangle  = v_2/\sqrt{2}$. In our convention the mass matrices are diagonalized as
\begin{equation}
\begin{split}
U^T M_U U_c &=  M_U^{\rm diag}, \qquad \, N^T M_\nu N_c = M_\nu^{diag},\\
D^T M_D D_c &= M_D^{\rm diag}, \qquad E^T M_E E_c = M_E^{\rm diag}.
\label{eq:masses}
\end{split}
\end{equation}
The neutrino masses are generated using the inverse seesaw mechanism (for more details see Ref.~\cite{FileviezPerez:2013zmv}).

\subsection{Flavor Violation and Fermion Masses}
In this theory the interactions mediating flavor violating processes for down quarks are proportional to the Yukawa matrix $Y_4$, which from Eq.~\eqref{eq:massesYuk} one can write as:
\begin{equation}
Y_4=\sqrt{\frac{3}{2}} \frac{(M_D - M_E)}{v \sin \beta}.
\label{eq:Y4}
\end{equation}
Here $v=\sqrt{v_1^2+v_2^2}=246$ GeV and $\tan \beta=v_2/v_1$. 
As one can appreciate, the Yukawa matrix $Y_4$ defines the difference between the mass matrices for down quarks and charged leptons. Therefore, the amount of flavor violation is bounded from above by the values of the quark masses in the flavor space.  A similar relation to Eq.~\eqref{eq:Y4} exists for $Y_2$. However, since the Dirac masses $M^D_\nu$ are unknown, the entries in $Y_2$ are not constrained and could even be zero. Contrarily, the entries of $Y_4$ are needed to correct the mass relation between the down quarks and charged leptons. Thus, in the following we will focus only on this matrix.

To understand the connection between the flavor structure of the SM and the flavor violation in this theory, let us consider the interactions of the scalar leptoquarks $\Phi_3$ and $\Phi_4$ mediated by the Yukawa matrix $Y_4$.\footnote{We could similarly proceed with the other scalars in $\Phi$, i.e. $H_2$ and $\Phi_8$. However, we focus on the scalar leptoquarks since we expect, motivated by the recent flavor anomalies, at least one of them to be at the TeV scale.} Starting from their components with electromagnetic charge $\pm \, 2/3$, 
\begin{eqnarray}
\bar{e}^i d^j \phi_3^{-2/3}&:& \hspace{0.5cm} i c_3^{ij} P_R = i \sqrt{\frac{3}{2}} \frac{1}{v \sin \beta}(V^\dagger M_D^{\rm diag}- M_E^{\rm diag} V_c^T)^{ij} P_R, \label{eq:FRphi3e}\\
 \bar{d}^i e^j \phi_4^{2/3}&:& \hspace{0.5cm} 
 i c_4^{ij}  P_R = i  \sqrt{\frac{3}{2}} \frac{1}{v \sin \beta}(M_D^{\rm diag} V_c^* - V M_E^{\rm diag})^{ij} P_R, \label{eq:FRphi4d}
 \end{eqnarray}
 where $P_{L,R}$ are the chiral projectors $P_{L,R} = (1 \mp \gamma_5)/2$, and 
the matrices $V$ and $V_c$ are unitary (mixing) matrices defined as $V=D^\dagger E$ and $V_c=D_c^\dagger E_c$, respectively. On the other hand, the Yukawa interactions mediated by $Y_4$ for their $\SU(2)_L$ partners are given by
\begin{eqnarray}
&&\bar{\nu}^i d^j \phi_3^{1/3}: \hspace{0.5cm} i \left( V_\textsc{pmns}^\dagger K_3^* \ c_3 \right)^{ij} P_R = i \sqrt{\frac{3}{2}} \frac{1}{v \sin \beta} (V_\textsc{pmns}^\dagger)^{ik}K_3^{*k} ( V^\dagger M_D^\text{diag} - M_E^\text{diag}V_c^T)^{kj}P_R, \label{eq:FRphi3nu} \\
&&\bar{u}^i e^j \phi_4^{5/3}: \hspace{0.5cm} i (K_1 V_\textsc{ckm} K_2 c_4)^{ij} P_R=  i \sqrt{\frac{3}{2}} \frac{1}{v \sin \beta} K_1^i V_\textsc{ckm}^{ik} K_2^k (M_D^{\rm diag} V_c^* - V M_E^{\rm diag})^{kj} P_R,\label{eq:FRphi4u}
\end{eqnarray}
where $K_1$, $K_2$ and $K_3$ are diagonal matrices containing phases. We note that, since the $V_\textsc{ckm}$ and $V_\textsc{pmns}$ have well-known structures, the above Feynman rules are also determined by the matrices $c_3$ and $c_4$. 
Notice that the matrices $c_3$ and $c_4$ define the interactions of the leptoquarks with the down quarks and charged leptons, and that the amount of flavor violation in these interactions is bounded by the correspondent fermion masses and mixings.
Neglecting the masses of the first generation, $m_e$ and $m_d$, we can write the matrices $c_3$ and $c_4$ as follows:
\begin{align}
c_3 &=  \sqrt{\frac{3}{2}} \frac{1}{v \sin \beta} \begin{pmatrix} 
 0 && m_s(V^*)^{21} && m_b(V^*)^{31} \\
 - m_\mu V_c^{12} && m_s(V^*)^{22} - m_\mu V_c^{22} &&  m_b(V^*)^{32}  - m_\mu V_c^{32}\\
 - m_\tau V_c^{13}   && m_s (V^*)^{23} - m_\tau V_c^{23} && m_b(V^*)^{33} - m_\tau V_c^{33}
 \end{pmatrix},  \label{equation:c3} \\
& \nonumber \\
c_4 &=  \sqrt{\frac{3}{2}} \frac{1}{v \sin \beta} \begin{pmatrix} 
 0 && - m_\mu V^{12} && - m_\tau V^{13} \\
 m_s (V_c^*)^{21} &&  m_s (V_c^*)^{22} - m_\mu V^{22}  && m_s (V_c^*)^{23} - m_\tau V^{23}   \\
 m_b (V_c^*)^{31}   && m_b (V_c^*)^{32} - m_\mu V^{32}  && m_b (V_c^*)^{33} - m_\tau V^{33} 
 \end{pmatrix}.
 \label{equation:c4}
\end{align}
Remarkably, the theory allows us to write the (in principle) arbitrary Yukawa matrix $Y_4$ in the physical basis for the fermions as a linear combination (L.C.) of two pieces: the known fermion masses ($M_f$) and two unitary matrices ($V_{(c)}$) which cannot be derived from first principles, i.e.
$Y_4 = \text{L.C.} \left[ M_f \,, V_{(c)} \right].$
Such factorization, as we will show in the upcoming sections, will allow us to determine the texture of $V$ and $V_c$ combining theoretical and experimental constraints, which will have strong implications on the quark-lepton unification scale as well as on establishing correlations amongst flavor observables. In the following we will assume
\begin{equation}
M_{\phi_3^{-2/3}} \sim M_{\phi_3^{1/3}} \sim M_{\Phi_3}, \qquad \text{ and } \qquad  M_{\phi_4^{2/3}} \sim M_{\phi_4^{5/3}} \sim M_{\Phi_4},
\label{eq:SU2splitting}
\end{equation}
for simplicity, since no large splitting is expected from the $\SU(2)_L$ corrections. Taking into account that perturbativity of the Yukawa couplings requires that $\sin \beta \gtrsim 0.01$, while direct searches at LHC demand $M_{\Phi_4}, M_{\Phi_3} \gtrsim 1 $ TeV~\cite{CMS:2018oaj,ATLAS:2020dsk,ATL-PHYS-PUB-2021-017,CMS:2020wzx}, the lower limit on the product $M_{\Phi_{3,4}} \sin \beta \gtrsim 10 \text{ GeV}$ must be fulfilled.

\section{Flavor Anomalies}
\label{sec:FlavorAnomalies}
%
Experimental measurements suggest violation of lepton flavor universality in processes involving $b \to s$ transitions, usually referred as {\it neutral anomalies}, and $b \to c$ transitions, known as {\it charged anomalies}. 
The theory for quark-lepton unification proposed in Ref.~\cite{FileviezPerez:2013zmv} predicts several fields beyond the SM, as we mentioned briefly in Sec.~\ref{sec:Theory}:

\begin{itemize}

\item Vector leptoquark, $X_\mu \sim (\mathbf{3},\mathbf{1},2/3)_\text{SM}$: This vector boson could explain the flavor anomalies and satisfy the strong experimental bound on $K_L \to e^{\pm} \mu^{\mp}$ amongst others only if we do not stick to the unitary constraints on the 
matrices defining the mixing between quarks and leptons, i.e. $V$ and $V_c$. Therefore, we do not see this solution very appealing even if it has been used often in the literature~\cite{Buttazzo:2017ixm,Calibbi:2017qbu,DiLuzio:2017vat,Assad:2017iib,Bordone:2017bld,Cornella:2019hct,Cornella:2021sby,Balaji:2019kwe}.

\item Scalar diquark, $\Phi_8 \sim (\mathbf{8},\mathbf{2},1/2)_\text{SM}$: The scalar diquark does not couple to leptons and then it cannot play a relevant role in the flavor anomalies.

\item New Higgs doublet, $H_2 \sim (\mathbf{1},\mathbf{2},1/2)_\text{SM}$: In this theory one has new physical Higgs bosons, the charged Higgs $H^{\pm}$, the heavy CP-even $H$ Higgs and the CP-odd $A$ Higgs. 
These Higgses cannot be used to address the anomalies because they provide only scalar operators~\cite{Ghosh:2017ber,Altmannshofer:2021qrr}.

\item Scalar leptoquark, $\Phi_3 \sim (\mathbf{\bar 3}, \mathbf{2}, 1/6)_\text{SM}$: The possibility of using $\Phi_3$ to explain the flavor anomalies was studied in Ref.~\cite{Perez:2021ddi}. 
Unfortunately, this solution requires large Wilson coefficients and one predicts generically too large branching ratios for the meson decays.

\item Scalar leptoquark, $\Phi_4 \sim (\mathbf{3},\mathbf{2},7/6)_\text{SM}$: This scalar leptoquark is the best candidate we have to address the flavor anomalies. Its couplings to down quarks and charged leptons are related to the unification of quarks and leptons. 
We have shown in the previous section that  the flavor violating interactions are bounded from above by the values of the quark and lepton masses. 

\end{itemize}

In this section we study in detail the possibility to explain the neutral flavor anomalies using the interactions of $\Phi_4$ with quarks and leptons. We relegate the study of the charged anomalies to Appendix~\ref{app:Charged} since they are not solely determined by the Yukawa matrix $Y_4$. Nevertheless, in that appendix we show how they can also be successfully explained by the leptoquark $\Phi_4$ by allowing for complex Wilson coefficients. The possibility of explaining both neutral and charged anomalies with this single scalar leptoquark was already noted in Ref.~\cite{Popov:2019tyc}, although in their study the textures are empirically adopted from the perspective of a simplified model. In our case, however, the minimal theory for quark-lepton unification will allow us to establish exact correlations amongst different flavor observables, as we will show in the upcoming sections.

The neutral anomalies are specially motivated after the latest measurement reported by the LHCb collaboration~\cite{LHCb:2021trn} on the ratio: 
\begin{equation}
{\cal R}_{K} \equiv \frac{\text{Br}(B \to K \mu^+ \mu^-) }{ \text{Br}(B \to K e^+ e^-)},
\label{eq:RK}
\end{equation} 
which deviates $3.1\, \sigma$ from the SM prediction. From the scalars predicted by the theory, the $\phi_4^{2/3}$ and $\phi_3^{-2/3}$ leptoquarks contribute to the processes involving $b \to s$ transitions through the effective interactions listed below,
\begin{equation}
-{\cal L}_\text{eff}^{b \to s} \supset
 \frac{(c_3^{li})^*c_3^{kj}}{2M_{\Phi_3}^2} (\bar d_R^i \gamma^\mu d_R^j) (\bar e_L^k \gamma_\mu e_L^l) + 
 \frac{ c_4^{il}(c_4^{jk})^*}{2M_{\Phi_4}^2}  (\bar d_L^i \gamma^\mu d_L^j)(\bar e_R^k \gamma_\mu e_R^\ell) + \text{h.c.}.
\end{equation}
Taking $i \! = \! 2$, $j \! = \! 3$, $l \! = \!  k \!  = \! \ell$, the above interactions can be identified with the following Wilson coefficients,
\begin{equation}
\begin{split}
C_{9\mu \mu}' = - C_{10\mu \mu}' &= \frac{\sqrt{2} \pi}{\alpha \, G_F V_\textsc{ckm}^{tb}(V_\textsc{ckm}^{ts})^*} \frac{(c_3^{22})^* c_3^{23}}{4 M_{\Phi_3}^2}, \quad C_{9 ee}'=-C_{10ee}' = \frac{\sqrt{2} \pi}{\alpha \, G_F V_\textsc{ckm}^{tb}(V_\textsc{ckm}^{ts})^*} \frac{(c_3^{12})^* c_3^{13}}{4 M_{\Phi_3}^2},\\
C_{9 \mu \mu} = C_{10 \mu \mu} &=\frac{\sqrt{2} \pi}{\alpha \, G_F V_\textsc{ckm}^{tb}(V_\textsc{ckm}^{ts})^*} \frac{c_4^{22} (c_4^{32})^*}{4 M_{\Phi_4}^2} ,  \quad C_{9 ee}=C_{10ee} =  \frac{\sqrt{2} \pi}{\alpha \, G_F V_\textsc{ckm}^{tb}(V_\textsc{ckm}^{ts})^*} \frac{c_4^{21} (c_4^{31})^*}{4 M_{\Phi_4}^2} ,
\end{split}
\label{eq:neutralWC0}
\end{equation}
in the context of the effective Lagrangian
\begin{equation}
\begin{split}
-{\cal L}_\text{eff}^{b \to s} \supset \frac{4 G_F}{\sqrt{2}} V_\textsc{ckm}^{tb} (V_\textsc{ckm}^{ts})^* \frac{\alpha}{4\pi} & \left[ C_{9 \ell \ell}(\bar s \gamma^\mu P_L b)(\bar \ell \gamma_\mu \ell) + C_{10 \ell \ell} (\bar s \gamma^\mu P_L b)(\bar \ell \gamma_\mu \gamma_5 \ell)  \right. \\ 
& \left. + C_{9 \ell \ell}' (\bar s \gamma^\mu P_R b) (\bar \ell \gamma_\mu \ell) + C_{10\ell \ell}' (\bar s \gamma^\mu P_R b)(\bar \ell \gamma_\mu \gamma_5 \ell) \right] + \text{h.c.}.
\end{split}
\end{equation}
Here, $G_F$ is the Fermi constant and $V_\textsc{ckm}^{tb}$ and $V_\textsc{ckm}^{ts}$ are elements of the Cabibbo-Kobayashi-Maskawa matrix. %
The anomalies we are interested to address appear in ${\cal R}_K$~\cite{LHCb:2021trn}, and ${\cal R}_{K^*}$~\cite{LHCb:2017avl} (the latter defined as Eq.~\eqref{eq:RK} but exchanging the $K$ by $K^*$):
\begin{equation}
\begin{split}
& {\cal R}_K^{\text{high-}q^2}= 0.846^{+0.042}_{-0.039} \, (\rm{stat.) \, }^{+ 0.013}_{-0.012} \, (\rm{syst.}), \\
& {\cal R}_{K^*}^{\text{low-}q^2} = 0.66^{+0.11}_{-0.07}\, (\rm{stat.})\pm 0.03 \, (\rm{syst.}),\\
& {\cal R}_{K^*}^{\text{high-}q^2} = 0.69^{+0.11}_{-0.07}\, (\rm{stat.}) \pm 0.05 \,(\rm{syst.}),
\label{eq:neutral}
\end{split}
\end{equation}
where high-$q^2$ refers to the integrated window $q^2 \subset [1.1,6] \text{ GeV}^2$ and low-$q^2$ to $q^2 \subset [0.045,1.1] \text{ GeV}^2$.
Recently, the LHCb collaboration reported~\cite{LHCb:2021lvy}
\begin{equation}
{\cal R}_{K^*} = 0.70^{+0.18}_{-0.13} \, (\rm{stat.) \,}^{+0.03}_{-0.04}\, (\rm{syst.}),
\end{equation}
in the $q^2$ range $[0.045,6]$ GeV$^2$. In the plots attached to this section we show the constraints coming from the ${\cal R}_{K^*}$ measurements in the different $q^2$ bins because they are more restrictive, although we have checked that the above measurement does not exclude any of the parameter space we show in the figures where there is overlap between the measurements in Eq.~\eqref{eq:neutral}.

\begin{figure}[t]
\includegraphics[width=0.46\linewidth]{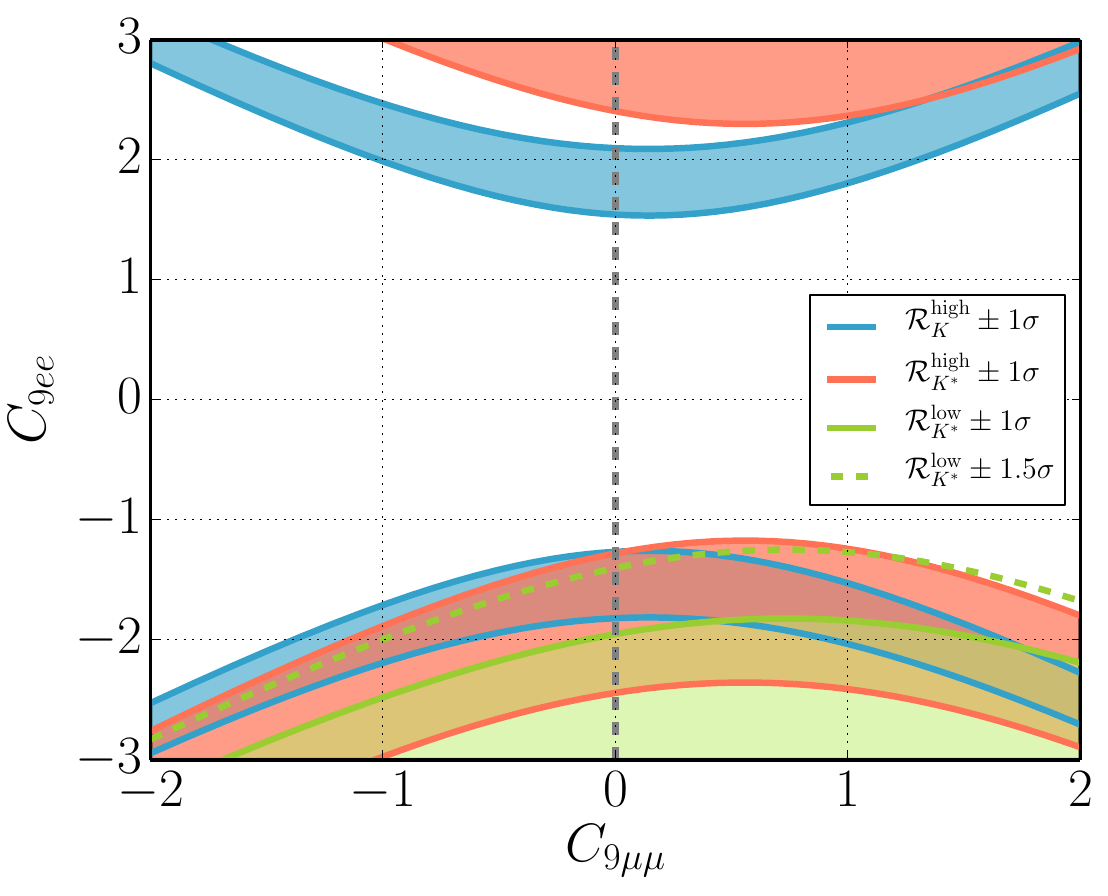}
\includegraphics[width=0.48\linewidth]{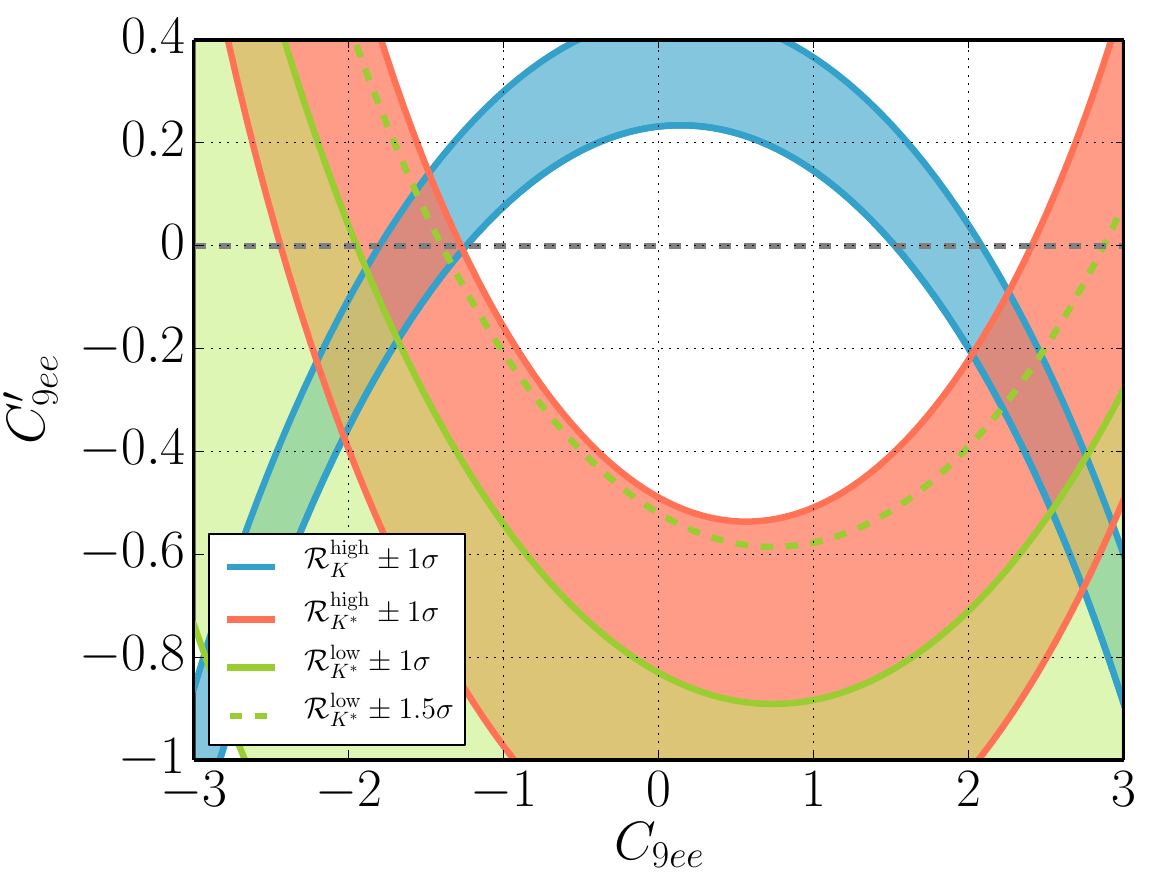}
\caption{Parameter space satisfying the different anomalous ratios ${\cal R}_{K^{(*)}}$ in Eq.~\eqref{eq:neutral} at $1 \,\sigma$ in the $C_{9\mu \mu}$ - $C_{9ee}$ (left panel) and $C_{9ee}$ - $C_{9ee}'$ (right panel) planes: in blue the ${\cal R}_K^\text{exp}$ within $1.1<q^2/\text{ GeV}^2<6$, in red ${\cal R}_{K^*}^\text{exp}$ for $1.1<q^2 / \text{ GeV}^2<6$, and in green ${\cal R}_{K^*}^\text{exp}$ for $0.045<q^2/\text{GeV}^2< 1.1$. The dashed line shows the ${\cal R}_{K^*}^\text{exp}$ for the low $q^2$ window at $1.5 \, \sigma$. Note that $C_{10ee} = C_{9ee}$, $C_{10 \mu \mu} = C_{9\mu \mu}$, and $C_{10ee}' = -C_{9ee}'$, according to Eq.~\eqref{eq:neutralWC0}.}
\label{fig:Neutral}
\end{figure}

We already mentioned that $\Phi_3$ alone cannot address the neutral anomalies in a consistent way~\cite{Perez:2021ddi}, however one could in principle consider the combined effect of both scalar leptoquarks $\Phi_3$ and $\Phi_4$. New physics affecting the muons is discouraged because (a) in the case of $\Phi_3$, the linear term in $C_{9\mu \mu}^{'} = -C_{10 \mu \mu}^{'}$ enters constructively and destructively in ${\cal R}_{K^*}$ and ${\cal R}_{K}$, respectively,\footnote{We refer the reader to Ref.~\cite{Perez:2021ddi} for the explicit formulae of the ratios as a function of the Wilson coefficients. } and hence a reduction of both ratios cannot be simultaneously achieved with small Wilson coefficients; (b) for $\Phi_4$, the quadratic term of $C_{9\mu\mu} = C_{10\mu\mu}$ in ${\cal R}_{K}$ is approximately $6$ times larger than the linear term, which rapidly dominates and precludes the possibility of reducing ${\cal R}_K$ with only the muonic interactions, hence requiring the help of large Wilson coefficients for the electrons, as illustrated in the left panel of Fig.~\ref{fig:Neutral}. Therefore, we proceed with a solution that only modifies the couplings to electrons. The latter is also motivated by the recent experimental measurement of $\text{Br}(B_s \to \mu^+ \mu^-)$ by CMS~\cite{CMS-PAS-BPH-21-006}, which is consistent with the SM prediction.

We note that, as the right panel in Fig.~\ref{fig:Neutral} shows, the tension amongst the ratios ${\cal R}_K$ and ${\cal R}_{K^*}$ could be explained (by relaxing ${\cal R}_{K^*}^{\text{low-}q^2}$ to $1.5 \, \sigma$) with the presence of only one Wilson coefficient: $C_{9ee} (=C_{10ee}) \simeq -1.4$. In Fig.~\ref{fig:Neutral}, the blue, red, and green shaded areas show where the experimental values of ${\cal R}_K^{\text{high-}q^2}$, ${\cal R}_{K^*}^{\text{high-}q^2}$, and ${\cal R}_{K^*}^{\text{low-}q^2}$ are met at $1\, \sigma$, respectively. We also show ${\cal R}_{K^*}^{\text{low-}q^2}$ at $1.5~ \sigma$ in a green dashed line.   
\begin{figure}[t]
\includegraphics[width=0.5\linewidth]{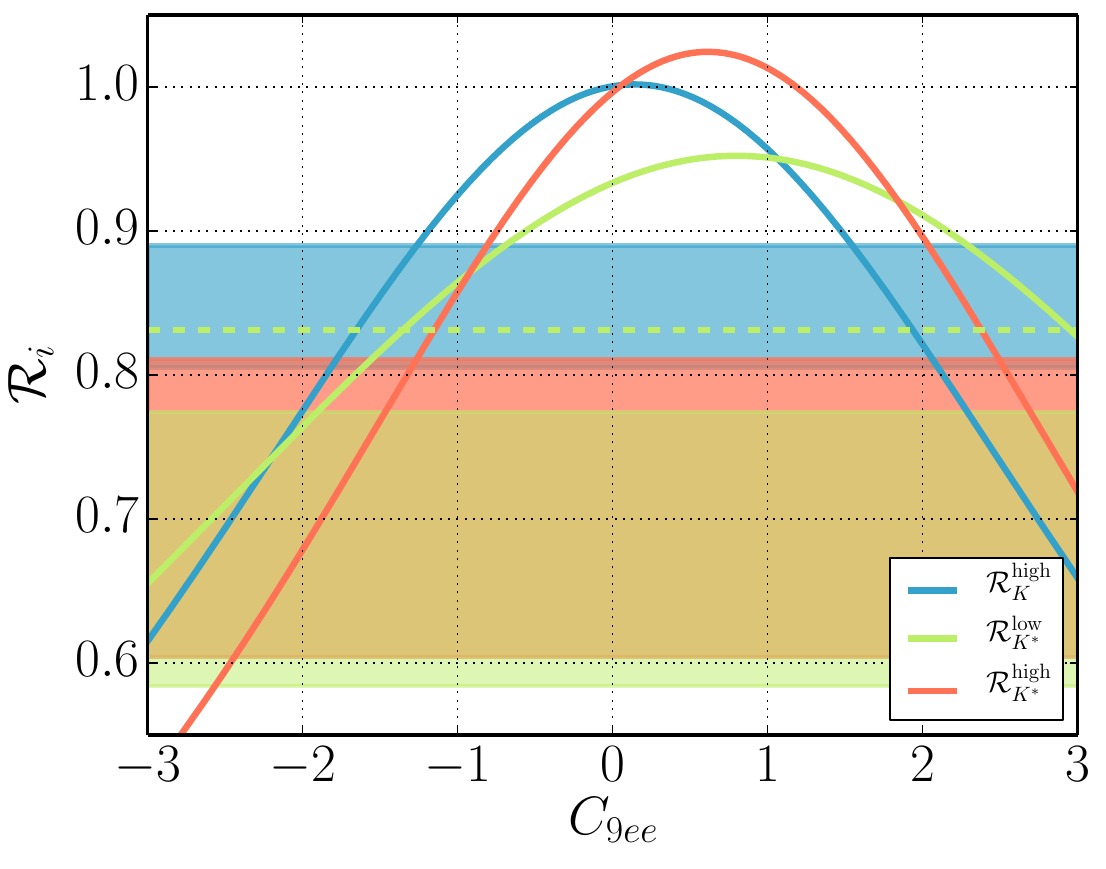}
\caption{Predictions for ${\cal R}_K$ integrated over $1.1<q^2/\text{ GeV}^2<6$ (blue), and ${\cal R}_{K^*}$ for the low $0.045<q^2/\text{GeV}^2< 1.1$ (green) and high $1.1<q^2/\text{ GeV}^2<6$ (red) bins, as a function of $C_{9ee}$. The shaded areas show where the respective experimental measurements are satisfied at $1~\sigma$. The green dashed line corresponds to the experimental constraint on the low-$q^2$ ${\cal R}_{K^*}$ at $1.5 \, \sigma$.}
\label{fig:C9e}
\end{figure}
In Fig.~\ref{fig:C9e} we show the predictions for the ratios ${\cal R}_{K^{(*)}}$ as a function of the relevant Wilson coefficient $C_{9ee}(=C_{10ee})$ for the neutral anomalies, which encodes the interaction between the leptoquark $\Phi_4$ and the electrons and quarks. The explicit dependence of the ratios on such Wilson coefficient are given by
\begin{equation}
\begin{split}
{\cal R}_K &= {\cal R}_K^\text{SM}  \left( 1 - 0.01781 \, \text{Re}\{C_{9ee}\} + 0.06359 |C_{9ee}|^2 \right)^{-1} \quad \ \text{ for } q^2\subset [1.1, \, 6] \, \text{ GeV}^2,\\
{\cal R}_{K^*} &= {\cal R}_{K^*}^\text{SM} \left(1 - 0.08886 \, \text{Re}\{ C_{9ee} \} + 0.07258 |C_{9ee}|^2\right)^{-1} \quad \ \text{ for } q^2\subset [1.1, \,6] \, \text{ GeV}^2,\\
{\cal R}_{K^*} &= {\cal R}_{K^*}^\text{SM} \left( 1 - 0.04912 \, \text{Re}\{C_{9ee}\} + 0.03078 |C_{9ee}|^2\right)^{-1} \quad \    \text{ for } q^2\subset [0.045, \, 1.1] \, \text{ GeV}^2,
\end{split}
\end{equation}
where the relevant Wilson coefficient is defined, in turn, as follows
\begin{equation}
C_{9 ee}(=C_{10ee}) =  \frac{3 \pi}{4\sqrt{2}\alpha \, G_F V_\textsc{ckm}^{tb}(V_\textsc{ckm}^{ts})^*} \frac{m_s \, m_b}{v^2} \frac{(V_c^*)^{21}V_c^{31}}{M_{\Phi_4}^2 \sin^2 \beta}.
\label{eq:WCPhi4}
\end{equation}
The above equation indicates which are the entries in the unitary mixing matrices required to address the neutral anomalies. 
As one can appreciate, this theory offers a scenario to explain the neutral anomalies using the fermion flavour violating interactions of the $\Phi_4$ leptoquark.
See the Appendix~\ref{app:Charged} for the explanation of the charged anomalies in the context of this leptoquark.

\section{Flavor Constraints}
\label{sec:MQL}
In the previous section we have discussed which elements of $V$ and $V_c$ are involved in explaining the observed deviations in the (expected) lepton flavor universal ratios ${\cal R}_{K^{(*)}}$. However, there are many other experimental bounds that will refine the shape of these matrices. For a TeV scale $\Phi_4$, particularly strong are the radiative leptonic decay $\mu \to e \gamma$ and the lepton flavor violating decay $K_L \to \mu^\pm e^\mp$, where $K_L$ stands for the long-lived neutral kaon. 

The main conditions that must be imposed to the textures of $V$ and $V_c$ are listed below:
\begin{enumerate}[(i)]
\item The neutral anomalies in the ${\cal R}_{K^{(*)}}$ ratios require $c_4^{21} \neq 0$ and $c_4^{31}\neq 0$, which translates into $V_c^{21} \neq 0$ and $V_c^{31} \neq 0$, respectively (see Sec.~\ref{sec:FlavorAnomalies}). Notice that, as Fig.~\ref{fig:C9e} shows, only the Wilson coefficient $C_{9ee} (= C_{10ee}) \neq 0$ is needed in order to address the experimental measurement of the high $q^2$ ratios ${\cal R}_{K^{(*)}}$ at $1 \, \sigma$ and ${\cal R}_{K^*}$ integrated between $ 0.045 < q^2 / \text{GeV}^2  < 1 $ at $1.5 \, \sigma$.

\item The radiative decays of charged leptons mediated by $\phi_4^{2/3}$ and $\phi_3^{-2/3}$ are not only chirality suppressed, but are also further suppressed by the ratio $(m_q / M_{\Phi_{3,4}})^4$ (where $m_q$ refers to the mass of the quark running inside the loop) because of a cancellation in the the loop functions due to the $\pm 2/3$ electric charge of the leptoquarks~\cite{Dorsner:2016wpm}. However, such cancellation does not apply to $\phi_4^{5/3}$, whose effect therefore dominates the contribution to the radiative leptonic decays through the diagrams shown in Fig.~\ref{fig:mutoeg}. 
%
%
\begin{figure}[t]
\includegraphics[width=0.7\linewidth]{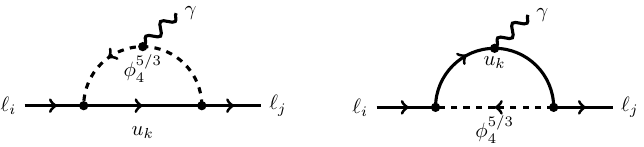}
\caption{Feynman diagrams for the topologies of the main contributions from the scalar leptoquarks to $\ell_i \to \ell_j \gamma$.}
\label{fig:mutoeg}
\end{figure}
Although being chirality suppressed, the stringent experimental constraint $\text{Br}(\mu \to e \gamma) < 4.2 \times 10^{-13}$~\cite{MEG:2016leq} sets the following bounds on the parameters of the theory:
\begin{eqnarray}
 &&   \sqrt{\left |(V_c^*)^{21} \left[(V^*)^{22}-\frac{m_s}{m_\mu} V_c^{22}  \right]  \right|}< 0.44 \left( \frac{M_{\Phi_4} |\sin \beta|}{10 \text{ GeV}} \right), \label{eq:mutoegPrimi}\\
 && \sqrt{\left | (V_c^*)^{31}\left[ V_c^{32} -\frac{m_\mu}{m_b}  (V^*)^{32} \right] \right|}   < 0.011  \left(\frac{M_{\Phi_4} |\sin \beta|}{10 \text{ GeV}}\right),
 \label{eq:mutoegCon}
\end{eqnarray}
where the bounds are set by requiring that the contribution of each quark, charm in Eq.~\eqref{eq:mutoegPrimi} and top in Eq.~\eqref{eq:mutoegCon}, saturates the experimental constraint.
While Eq.~\eqref{eq:mutoegPrimi} could be easily satisfied, the condition in Eq.~\eqref{eq:mutoegCon} requires $ V_c^{32} - (m_\mu / m_b) (V^*)^{32}  \simeq 0 $, i.e. the texture of matrix $c_4$, displayed in Eq.~\eqref{equation:c4}, must contain a zero in the element $c_4^{32}$. We note that $V_c^{31}$ is needed to address the neutral anomalies and therefore cannot be suppressed.  
\item Similarly, the strongest bound from Kaon decays, $\text{Br}(K_L \to e^\pm \mu^\mp) < 4.7 \times 10^{-12}$~\cite{Zyla:2020zbs}, requires
\begin{equation}
\sqrt{\left | V^{21}V_c^{12} \right| } \left(\frac{10 \text{ GeV}}{M_{\Phi_3} \sin \beta}\right) \ \text{ and } \ \sqrt{\left | V^{12}V_c^{21} \right|} \left(\frac{10 \text{ GeV}}{M_{\Phi_4} \sin \beta}\right) < 0.10.
\label{eq:Klong}
\end{equation}
While the constraint acting on the contribution coming from $\Phi_3$ can be easily avoided by having a heavy $\Phi_3$, the bound on the $\Phi_4$ leptoquark translates into $V^{12} \to 0$ because, similarly to the previous case, a non-zero $V_c^{21}$ is needed to address the neutral anomalies.
\end{enumerate}
Taking the above conditions into account, let us now exploit the fact that the matrices $V$ and $V_c$ are unitary to shape their textures.
Let us start from a generic parametrization of a unitary 3 x 3 matrix where we have taken the imaginary phases to be zero for simplicity
\footnote{The constraints from electric dipole moments (EDMs) give us a non-trivial bound on these phases. In this theory one has one-loop contributions to, for instance, the electron EDM with the leptoquark $\Phi_4$ inside the loop, the Higgs doublets $H_1$ and $H_2$ changing the quark chirality inside the loop, while the electron chirality is changed in the external line. We will investigate the EDM constraints in a future publication.}
\begin{equation}
\begin{pmatrix} c_{12} c_{13} & s_{12} c_{13} & s_{13} \\ - s_{12} c_{23} - c_{12} s_{23} s_{13} & c_{12} c_{23} - s_{12} s_{23} s_{13} & s_{23} c_{13} \\ s_{12} s_{23} - c_{12} c_{23} s_{13} & -c_{12} s_{23} - s_{12} c_{23} s_{13}  & c_{23} c_{13} \end{pmatrix},
\label{eq:unitarity}
\end{equation}
where $c_{ij} \equiv \cos \theta_{ij}$ and $s_{ij} \equiv \sin \theta_{ij}$, and $\theta_{12}$, $\theta_{13}$ and $\theta_{23}$ are Euler angles.

Starting from $V_c$, from condition (i) we know that the elements $V_c^{21}$ and $V_c^{31}$ cannot be suppressed. Condition (ii), however, requires $V_c^{32} \simeq (m_\mu/m_b) V^{32}$. Unitarity demands then that $V^{32}_c \lesssim 0.1$. The only possibility that consistently satisfies conditions (i) and (ii) is  $c_{12} \to \epsilon$ and $s_{13}\to \epsilon'$, being $\epsilon$ and $\epsilon'$ small parameters according to conditions (i) and (ii).\footnote{There is freedom in choosing the sign of the complementary trigonometric function of the one we take to be small. We will assume it positive without loss of generality.} 
Hence,
\begin{equation}
V_c = \begin{pmatrix} \epsilon & 1 & \epsilon' \\ -\cos \theta_c & \epsilon \cos \theta_c - \epsilon' \sin \theta_c & \sin \theta_c \\  \sin \theta_c &-\epsilon \sin \theta_c + \epsilon' \cos \theta_c & \cos \theta_c \end{pmatrix},
\label{eq:Vctexture}
\end{equation}
where we have neglected order ${\cal O}(\epsilon^2), {\cal O}(\epsilon'^2)$ and ${\cal O}(\epsilon \epsilon')$ terms. 

For the matrix $V$, on the other hand, there is more freedom. Condition (iii) requires that $V^{12} \to 0$.  We will assume without loss of generality that $s_{12} \to 0$,\footnote{Note that the other possibility $c_{13}\to 0$ is already included in the left-hand-side of Eq.~\eqref{eq:Vtexture} if one indeed takes $c_{13} \to 0$.}
\begin{equation}
 V= \begin{pmatrix} c_{13} && 0 && s_{13} \\ -s_{23} s_{13} && c_{23} && s_{23}c_{13} \\ -c_{23}s_{13} && -s_{23} && c_{23}c_{13} \end{pmatrix} \stackrel{\phantom{\frac{A}{A}}\displaystyle c_{23}\to \epsilon'' \phantom{\frac{A}{A}}}{=} \begin{pmatrix} \cos \theta && 0 && \sin \theta \\  -\sin \theta && \epsilon'' && \cos \theta \\ -\epsilon'' \sin \theta && -1 && \epsilon'' \cos \theta \end{pmatrix},
 \label{eq:Vtexture}
\end{equation}
where in the last equality we have taken the limit of $c_{23} \to \epsilon''$ to be consistent with the stringent experimental bounds on lepton flavor violating processes involving muons, and we have relabelled the remaining angle $c_{13}$ as $\cos \theta$.
Using the textures for $V$ and $V_c$, the Yukawa matrix for the $\Phi_4$ leptoquark reads
\begin{eqnarray}
c_4 &= &  \frac{\sqrt{3/2}}{v \sin \beta}  \begin{pmatrix} 
 0 && 0 && -m_\tau \sin \theta  \\
-m_s \cos \theta_c && m_s (\epsilon \cos \theta_c - \epsilon' \sin \theta_c) - m_\mu \epsilon''  &&  m_s \sin \theta_c - m_\tau \cos \theta \\
 m_b \sin \theta_c  && 0 && m_b \cos \theta_c  
\end{pmatrix}, \nonumber \\
&\simeq &
\frac{\sqrt{3/2}}{v \sin \beta} \begin{pmatrix} 
 0 && 0 && -m_\tau \sin \theta  \\
-m_s \cos \theta_c && 0 &&  m_s \sin \theta_c - m_\tau \cos \theta \\
 m_b \sin \theta_c  && 0 && m_b \cos \theta_c  
\end{pmatrix}.
 \label{eq:c4texture}
\end{eqnarray}
Notice that, in the limit $\epsilon'' \to 0$, condition (ii) fixes $(\epsilon \sin \theta_c + \epsilon' \cos \theta_c) \simeq -m_\mu/m_b$. Hence, the element $c_4^{22} $ is expected to be of the order of the first generation of quark masses and therefore we will neglect it in the following.

The above matrix defines the Yukawa interactions of $\phi_4^{2/3}$ with the SM fermions, as parametrized in Eq.~\eqref{eq:FRphi3e}. The Feynman rules of its $\SU(2)_L$ partner, $\phi_4^{5/3}$, are also determined by the texture above, as Eq.~\eqref{eq:FRphi4u} shows, up to known mixing matrices. For the latter case we will adopt $K_1 V_\textsc{ckm} K_2 c_4 \sim c_4$.
It worths to emphasize that knowing the textures of the matrices $V$ and $V_c$ allows us to predict the interactions of the other scalars in the $\Phi \sim (\mathbf{15},\mathbf{2},1/2)$ representation, i.e. the scalar leptoquark $\Phi_3$, the diquark $\Phi_8$ and the second Higgs $H_2$. However, since only a light (TeV scale) $\Phi_4$ is required for consistency with current experimental data, the rest of the scalar fields in $\Phi$ could be heavier as long as the following sum rule coming from the scalar potential is respected~\cite{Faber:2018qon},
\begin{equation}
M_{\Phi_8}^2 + 2 M_{H_2}^2 = \frac{3}{2}\left(M_{\Phi_3}^2 + M_{\Phi_4}^2\right),
\end{equation}
and hence easily avoid the current bounds. 

Now, applying the texture of $c_4$ from Eq.~\eqref{eq:c4texture} in Eq.~\eqref{eq:WCPhi4}, the Wilson coefficient contributing to the anomalies is given by 
\begin{equation}
 C_{9 ee}(=C_{10ee}) = - \frac{3\sqrt{2} \pi}{16 \, \alpha G_F V_\textsc{ckm}^{tb}(V_\textsc{ckm}^{ts})^*}  \frac{ m_s  m_b}{v^2 }\frac{\sin 2 \theta_c}{M_{\Phi_4}^2 \sin^2 \beta}.
 \end{equation}
The experimental requirement of $C_{9ee}(=C_{10ee})\sim -1.4$ therefore translates into the following condition,
\begin{equation}
\frac{\sin 2 \theta_c}{\sin^2 \beta \, M_{\Phi_4}^2} \simeq \frac{1}{1174  \text{ GeV}^2},
\label{eq:conditionangles}
\end{equation}
where $\sin 2\theta_c \gtrsim 0.085$ to be consistent with perturbativity of the Yukawa couplings and collider bounds.

\section{Flavor signatures}
\label{sec:FlavorSignatures}
In this section we study the consequences of adopting the texture in Eq.~\eqref{eq:c4texture} for the couplings in $c_4$ (i.e. interactions of $\Phi_4$ with SM fermions). 

\begin{figure}[t]
\includegraphics[width=0.6\linewidth]{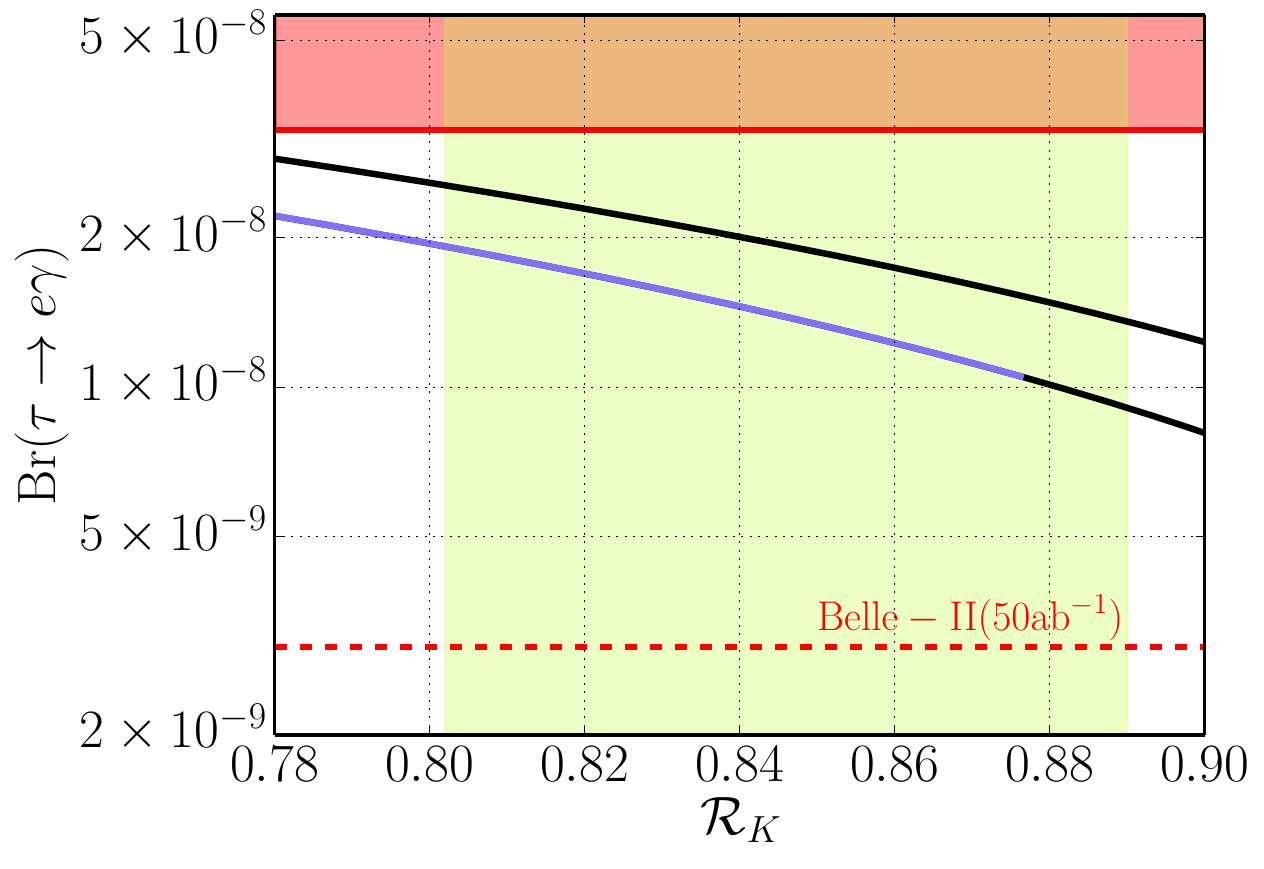}
\caption{Predictions for the branching ratio $\tau \to e \gamma$ and ${\cal R}_K$. The black line shows the correlation between these two observables, while the purple line corresponds to the values of $C_{9ee}$ that are in agreement with the ${\cal R}_{K^*}$ as required in the previous sections. The red line corresponds to the current bound $\text{Br} (\tau \to e \gamma ) < 3.3 \times 10^{-8}$~\cite{BaBar:2009hkt} from BaBar, while the dashed line shows the projected bound by the Belle-II collaboration~\cite{Belle-II:2018jsg}.}
\label{fig:tauegamma}
\end{figure}

\begin{itemize}

\item $ {\ell_i \to \ell_j \gamma}$:
Radiative decays of charged leptons can impose severe constraints on lepton flavor violating processes. The branching ratio of a charged lepton decaying into a lighter charged lepton and a photon in this theory is given by,
\begin{equation}
\text{Br}( \ell_i \to \ell_j \gamma ) \simeq \tau_{\ell_i} \frac{\alpha}{4} m_{\ell_i}^5 \left | \frac{3}{64\pi^2M_{\Phi_4}^2}  \sum_{q=u,c,t} c_4^{q i} (c_4^{q j})^* \right|^2  ,
\label{eq:mutoeg}
\end{equation}
where $\tau_{\ell_i}$ is the lifetime of the lepton $\ell_i$, and $\alpha$ is the fine-structure constant. The mediator in Eq.~\eqref{eq:mutoeg} is the leptoquark $\phi_4^{5/3}$, which will be the only leptoquark giving a relevant contribution to these decays as discussed earlier. The texture adopted for the Yukawa interactions of $\Phi_4$ avoids (as demanded) the strong bound $\text{Br} (\mu \to e \gamma ) < 4.2 \times 10^{-13}$~\cite{MEG:2016leq}. However, it can mediate the decay channel $\tau \to e \gamma$ with a contribution proportional to $|c_4^{23}(c_4^{21})^* + c_4^{33}(c_4^{31})^*|^2$. Since the contribution mediated by the top quark, i.e. $c_4^{33}(c_4^{31})^*$ involves two powers of the bottom mass, it will dominate the decay. Hence, applying the texture in Eq.~\eqref{eq:c4texture}, we obtain
\begin{equation}
\text{Br}(\tau \to e \gamma) \simeq \tau_\tau \frac{\alpha}{4} m_\tau^5 \left(\frac{3}{64\pi^2}\right)^2 \left( \frac{3}{2}\right)^2\frac{m_b^4}{4 v^4}  \left| \frac{\sin 2 \theta_c}{M_{\Phi_4}^2 \sin^2 \beta}\right|^2 \simeq 1.1 \times 10^{-8},
\end{equation}
where in the last equation we have applied the condition in Eq.~\eqref{eq:conditionangles}. The above prediction lies very close to the current experimental bound $\text{Br}(\tau \to e \gamma) < 3.3 \times 10^{-8}$~\cite{BaBar:2009hkt}, only away by a factor of three. The projected limits on these radiative decays are expected to improve in an order of magnitude~\cite{Aushev:2010bq,Baldini:2013ke}, which will probe the potential of $\Phi_4$ to address the ratios ${\cal R}_{K^{(*)}}$, as we show in Fig.~\ref{fig:tauegamma}. We note that the fixed correlation between the ratios ${\cal R}_{K^{(*)}}$ and the radiative decay $\tau \to e \gamma$ is a consequence of having a UV completion behind the leptoquark interactions.
%
\item {\bf $  \tau \to \ell_i  \bar \ell_j  \ell_k $}:
%
The experimental bounds on the branching fraction of the tau decay to three charged leptons are of the order $\sim 10^{-8}$~\cite{Zyla:2020zbs}. Both penguin and box diagrams can contribute to the leptonic tau decays:
\begin{equation}
\begin{gathered}
\includegraphics[width=0.75\linewidth]{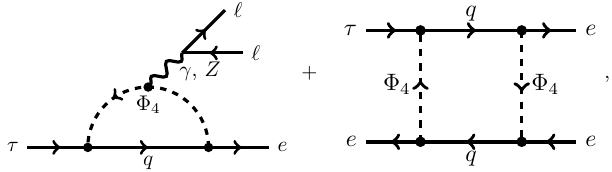}
\end{gathered}
\end{equation}

\vspace{-0.3cm}

where the quark (or quarks) mediating the interaction are down quarks if $\phi_4^{2/3}$ mediates the quantum process, or up quarks if it is $\phi_4^{5/3}$ instead. Note that chiral enhanced interactions are not present since there is only the leptoquark interaction with the right-handed charged lepton. The muonic channels $\tau^- \to \mu^- \mu^+ \mu^-$,  $\tau^- \to \mu^- e^+ e^-$, $\tau^- \to e^+ \mu^+ \mu^-$, and $\tau^- \to \mu^+ e^+ e^-$ cannot occur in the context of the $c_4$ texture in Eq.~\eqref{eq:c4texture}, however, note that $\Phi_4$ can still contribute to the channel $\tau^- \to e^- \mu^+ \mu^-$ through the penguin diagram on the left-hand-side of Eq.~\eqref{eq:penguins}. 

The channels $\text{Br}(\tau^- \to e^- \mu^+ \mu^-), \text{Br}(\tau^- \to e^- e^+ e^-)< 2.7 \times 10^{-8}$~\cite{Zyla:2020zbs} pick a contribution from $\Phi_4$ proportional to $|c_4^{23}(c_4^{21})^* + c_4^{33}(c_4^{31})^*|^2$, where we can neglect the first term in the sum in front of $c_4^{33}(c_4^{31})^*$ since, as in the pervious case, the latter is proportional to two powers of the bottom quark mass. Note that such combination of couplings entering in the decay is totally fixed by Eq.~\eqref{eq:conditionangles}. Therefore, we can estimate the expected branching fractions mediated by $\Phi_4$, $\text{Br}(\tau^- \to e^- \mu^+ \mu^-) \simeq {\cal O}(10^{-10})$ and $\text{Br}(\tau^- \to e^- e^+e^-) \simeq {\cal O}(10^{-10})$, where the exact value will be given by the precise value of the leptoquark mass which enters normalized by the quark masses logarithmically in the rates.\footnote{For the ${\cal O}(10^{-10})$ predictions of the leptonic branching fractions we have considered the leptoquark $\Phi_4$ to be at the ${\cal O}(1)$ TeV scale.}

\item {\textit{Meson mixing:}}
The contribution from a leptoquark with chiral couplings to the mixing between neutral mesons $M_{q_i \, \bar q_j}$ is given by the following effective Hamiltonian,
\begin{equation}
H_{\Delta F=2}^{\Phi_4} \simeq \frac{\left (\sum_\ell (c_4^{q_i \ell})^* c_4^{q_j \ell} \right)^2}{128\pi^2 M_{\Phi_4}^2} \left( \bar q_i \gamma_\mu P_{R} \, q_j \right ) (\bar q_i \gamma^\mu P_{R} \, q_j),
\end{equation}
where we have used that the limit of the loop function is 1 for suppressed ratios $m_\ell / M_{\Phi_4}$.
We list below the dependence of the meson mixing on the Yukawa interactions in $c_4$. 
\begin{equation*}
\begin{gathered}
\includegraphics[width=0.5\linewidth]{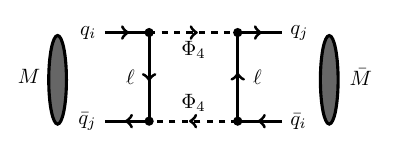}
\end{gathered}
\quad
\begin{cases}
K - \bar K \ \text{ mixing}: &  c_4^{2 3} (c_4^{1 3})^*\\
D - \bar D \ \text{ mixing}: &  c_4^{2 3} (c_4^{1 3})^*\\
B - \bar B \ \ \text{  mixing}: & c_4^{33}(c_4^{13})^*\\
B_s - \bar B_s \text{ mixing}: & c_4^{31}(c_4^{21})^* + c_4^{33}(c_4^{23})^*
\end{cases}
\end{equation*}
For $K-\bar K$ mixing, because the SM prediction is sensitive to long-distance contributions which are hard to quantify, we adopt the conservative approach that the the contribution of $\phi_4^{2/3}$ saturates the kaon mass difference~\cite{Mandal:2019gff}, $\Delta m_K = (3.484 \pm 0.006) \times 10^{-15}$ GeV~\cite{Zyla:2020zbs}, which translates into the following bounds:
\begin{equation}
M_{\Phi_4}\lesssim 15 \text{ TeV} \left|\frac{\cos \theta_c}{\sin \theta}\right|  , \qquad \text{ and } \qquad  M_{\Phi_4} \lesssim 0.81 \text{ TeV} \left| \frac{\sin 2 \theta_c}{\sin 2 \theta}\right|,
\label{eq:boundsKKbar}
\end{equation}
where we have assumed no cancellations between the two different terms in the entry $c_4^{23}$.
The upper bound on the leptoquark mass arises as a consequence of having the product $(\sin \beta M_{\Phi_4})$ constrained by the neutral anomalies, see Eq~\eqref{eq:conditionangles}. The above constraints are relaxed in the limit of a small $\theta$.

The $D-\bar D$ mixing will constrain the same combinations of $c_4$ entries. However, because the experimental bounds are weaker~\cite{Zyla:2020zbs} than those for the kaon mixing, they are automatically satisfied in the context of the bounds from Eq.~\eqref{eq:boundsKKbar}.

For the $B_{(s)} -\bar B_{(s)}$ mixing, the mass difference is given by~\cite{Saha:2010vw,Becirevic:2015asa}
\begin{equation}
\Delta m_{B_{ q}}= \Delta m_{B_{q}}^\text{SM} \left | 1 - \frac{S_0^{-1}(x_t)}{ \left| V_{tb}V_{tq}^*\right |^2} \frac{|\sum_\ell c_4^{3\ell}(c_4^{q \ell})^*|^2}{32 G_F^2m_W^2 M_{\Phi_4}^2}\right | \ \text{ with } \ S_0(x) = \frac{4x \! -\! 11x^2 \! +\!  x^3}{4(1-x)^2}-\frac{3x^3\ln x}{2(1-x)^3},
\end{equation}
where $x_t = m_t^2/m_W^2$. We will require that the new physics contribution is below $10\%$ of the SM one, where the rough assumption of $10\%$ error comes from the lattice calculations of the hadronic matrix elements~\cite{HFLAV:2019otj}. The constraint from $B-\bar B$ mixing, 
\begin{equation}
M_{\Phi_4} \lesssim 0.27 \text{ TeV} \left|\frac{\sin \theta_c}{\sin \theta}\right|,
\end{equation}
can be evaded, as in the previous $K-\bar K$ case, in the limit where $\theta$ is small. On the other hand, $B_s - \bar B_s$ mixing impose the following upper bounds on the leptoquark mass,
\begin{equation}
M_{\Phi_4} \lesssim 24.87 \text{ TeV}, \qquad \text{ and } \qquad M_{\Phi_4} \lesssim 1.34 \text{ TeV} \left | \frac{\sin \theta_c}{\cos \theta}\right | .
\label{eq:BsBsmix}
\end{equation}
where we have similarly required that the new physics contribution is below $10\%$ of the SM contribution. The constraint on the left-hand-side of Eq.~\eqref{eq:BsBsmix} comes from the contribution of $c_4^{31}(c_4^{23})^*$, while the bound on the right-hand-side comes from $c_4^{33}(c_4^{23})^*$, which clearly dominates. We note that, in order to be in agreement with the current experimental constraints on the mass of $\Phi_4$ from direct searches~\cite{CMS:2018oaj,ATLAS:2020dsk,ATL-PHYS-PUB-2021-017,CMS:2020wzx}, $\theta_c$ cannot be small, particularly $\sin \theta_c \gtrsim 0.7$. 

The above constraints suggest $\sin \theta$ to be small. In the following we will work in the limit $\theta \ll \theta_c$ and therefore neglect the channels involving entries proportional to $\sin \theta$. 

\item {\textit{Hadronic $ \tau$  decays:}}
%
The tau decays to hadrons are rare processes in the SM and suffer from experimental bounds that are order $10^{-7}-10^{-8}$ on their branching fractions.
The branching ratio of the $\tau$ decay to a pseudoscalar and charged lepton is given by
\begin{equation}
    \text{Br}(\tau \to P_{\bar ij} \, \ell_k) \simeq \tau_\tau \frac{f_P^2}{128\pi} \frac{(m_\tau^2-m_P^2)^2}{m_\tau} \left | \frac{c_4^{i3}(c_4^{j k})^*}{M_{\Phi_4}^2}  \right |^2 ,
\end{equation}
while the $\tau$ decay to a vector meson and charged lepton is given by the above expression but substituting $f_P^2(m_\tau^2-m_P^2)^2 \to f_V^2 (m_\tau^2-m_V^2)(m_\tau^4+m_V^2m_\tau^2-2m_V^4)/m_\tau^2$. 
In Table~\ref{tab:taudecays} we list the different decays that have a non-zero contribution in the context of the texture in Eq.~\eqref{eq:c4texture} and the limit $\sin \theta \to 0$. In there, we write the predictions of the theory as a function of the unknown parameters. In the cases where, for certain values of the parameters, the prediction may be in tension with the experimental bound, we explicitly show how they are restricted. We will follow the same approach in the upcoming subsections when showing the impact of the bounds in tables.

\begin{table}[h]
\begin{tabular}{|c | c | c | c|}
\hline
Channel & Non-zero contributions & Exp. bound~\cite{Zyla:2020zbs} & Prediction / Constraint \\
\hline
\hline
$\text{Br}(\tau \to \eta \, e)$ & $\propto |c_4^{23}(c_4^{21})^*|^2$ & $< 9.2 \times 10^{-8}$ &  $|\sin \theta_c | \gtrsim 0.042$ \\
$\text{Br}(\tau \to \eta' \, e)$ & $\propto |c_4^{23}(c_4^{21})^*|^2$ & $ <1.6 \times 10^{-7}$ &  $\simeq 6.7 \times 10^{-11} \left (\sin \theta_c\right)^{-2}$ \\
\hline
$\text{Br}(\tau \to \phi \, e)$ & $\propto |c_4^{23}(c_4^{21})^*|^2$ & $ < 3.1 \times 10^{-8}$ & $ |\sin \theta_c| \gtrsim 0.16$ \\
\hline
\end{tabular}
\caption{Hadronic decays of the $\tau$ mediated by the scalar leptoquark $\Phi_4$.}\label{tab:taudecays}
\end{table}
%

\item {\textit{Meson leptonic decays:}}
%
The leptonic decays of a meson are given by
\begin{equation}
\begin{split}
\text{Br}(M[q_j \bar q_i] \to \ell_1^- \ell_2^+) &= \frac{\tau_M}{256\pi}\frac{f_M^2}{m_M^3}\lambda^{1/2}(m_M^2,m_{\ell_1}^2,m_{\ell_2}^2) \left| \frac{c_4^{i\ell_2}(c_4^{j\ell_1})^*}{M_{\Phi_4}^2}\right|^2 \times \\
&\left ( m_M^2(m_{\ell_1}^2+m_{\ell_2}^2)-(m_{\ell_1}^2-m_{\ell_2}^2)^2 \right),
\end{split}
\end{equation}
where $\lambda(a,b,c) = a^2 + b^2+c^2 -2ab-2bc-2ca$ is the K\"{a}ll\'en function. For the channels that involve equal leptons in the final state, one should also include the contributions from the SM ($C_9^\text{SM} \simeq 4.211$, $C_{10}^\text{SM} =-4.103$~\cite{Blake:2016olu}). In Table~\ref{tab:mesonLeptonic} we list the contributions of the scalar leptoquark $\Phi_4$ to the leptonic decays of the mesons according to the $c_4$ texture in Eq.~\eqref{eq:c4texture} and the limit $\sin \theta \to 0$.
\begin{table}[h]
\begin{tabular}{|c | c | c | c|}
\hline
Channel & Non-zero contributions & Exp. bound~\cite{Zyla:2020zbs} & Prediction / Constraint \\
\hline
\hline
$\text{Br}(B_s \to e^+ e^-)$ & $\propto |c_4^{21}(c_4^{31})^*|^2 $ & $ < 2.8 \times 10^{-7}$ & $\simeq 1.2 \times 10^{-13}$ \\
\hline
$\text{Br} (B_s \to e^+ \tau^-)$ & $\propto |c_4^{31}(c_4^{23})^*|^2$ & $-$ & $\simeq 1.3 \times 10^{-5} \left(\cos \theta_c\right)^{-2}$ \\
\hline
$\text{Br} (B_s \to \tau^+ e^-)$ & $\propto |c_4^{33}(c_4^{21})^*|^2$ & $-$ & $ \ \ \simeq 3.7 \times 10^{-8} \left(\tan \theta_c \right)^{-2}$ \\
\hline
$\text{Br}(B_s \to \tau^+ \tau^-)$ & $\propto |c_4^{33}(c_4^{23})^*|^2$ & $<6.8 \times 10^{-3}$ & $ |\sin \theta_c| \gtrsim 0.06$ \\
\hline
\end{tabular}
\caption{Leptonic decays of mesons mediated by the scalar leptoquark $\Phi_4$. For the decay channel $B_s \to e^+ e^-$ we have included the SM contribution, while for the decay $B_s \to \tau^+ \tau^-$, the bound have been obtained by requiring the new physics to saturate the experimental bound.}\label{tab:mesonLeptonic}
\end{table}

\item {\textit{Charged semileptonic decays of mesons:}}
\begin{table}[t]
\begin{tabular}{ | c | c  | c  | c | c  | c | c |}
\hline
Coefficient & $e^+ e^- \ [c_4^{21} (c_4^{31})^*$ & \multicolumn{2}{c|}{$e^+ \tau^- \ [c_4^{31}(c_4^{23})^*]$ and $\tau^+ e^- \ [c_4^{33}(c_4^{21})^*]$} & $\tau^+\tau^- \ [c_4^{33}(c_4^{23})^*]$ \\
\hline
\hline
 &   $q^2 \subset [1.1,6]$ & full $q^2$ range & $q^2 \subset [(m_e+m_\tau)^2,6]$ & full $q^2$ range \\
 \hline
  $a_{B \to K \ell_1 \ell_2}$   & $\phantom{-}1.43 \times 10^{-7}$ & 0 & 0 & $\phantom{-}1.29 \times 10^{-7}$ \\
    $b_{B \to K \ell_1 \ell_2}$   &  $-2.56 \times 10^{-9}$ & 0 & 0 & $-2.47 \times 10^{-8}$ \\
      $c_{B \to K \ell_1 \ell_2}$   &  $\phantom{-}9.13 \times 10^{-9}$ & $1.96 \times 10^{-8}$ & $1.22 \times 10^{-9}$ & $\phantom{-}8.10 \times 10^{-9}$ \\
    \hline
  $a_{B \to K^* \ell_1 \ell_2}$ &  $\phantom{-}4.74 \times 10^{-6}$ & 0 & 0 & $2.43 \times 10^{-6}$ \\
    $b_{B \to K^* \ell_1 \ell_2}$ &  $-4.21  \times 10^{-7}$ & 0 & 0 & $5.96 \times 10^{-7}$ \\
        $c_{B \to K^* \ell_1 \ell_2}$ &  $\phantom{-}3.44 \times 10^{-7}$ & $7.65 \times 10^{-7}$ & $4.19 \times 10^{-8}$ & $1.79 \times 10^{-7}$ \\
    \hline
  $a_{B_s \to \phi \ell_1 \ell_2}$  &   $\phantom{-}5.11 \times 10^{-6}$ & $0$ & $0$ & $2.27\times 10^{-6}$ \\
    $b_{B_s \to \phi \ell_1 \ell_2}$  &   $-4.67 \times 10^{-7}$ & 0 & 0 & $5.98 \times 10^{-7}$ \\
      $c_{B_s \to \phi \ell_1 \ell_2}$  &   $\phantom{-}3.73 \times 10^{-7}$  & $7.74 \times 10^{-7}$ & $4.44 \times 10^{-8}$ & $1.70 \times 10^{-7}$ \\
    \hline
\end{tabular}
\caption{Semileptonic decays mediated by $\Phi_4$. Coefficients for the channels $B \to K^{(*)}$ and $B_s \to \phi$ involving different final leptons and $q^2$ integration ranges. The full $q^2$ range goes from $q^2_\text{min} = (m_{\ell_1}+m_{\ell_2})^2$ to $q^2_\text{max} = (m_{B_{(s)}}-m_{M^{(*)}})^2$.}
\label{tab:mesondecays}
\end{table}
The leptoquark $\Phi_4$ can contribute to the leptonic decays of mesons, as we know from the discussion in Sec.~\ref{sec:FlavorAnomalies} when studying its effect on the ratios ${\cal R}_{K^{(*)}}$. From the $c_4$ texture in Eq.~\eqref{eq:c4texture}, in the limit $\sin \theta \to 0$ (as the bounds on meson mixing require), only the decays $B \to K^{(*)} + \text{leptons}$ and $B_s \to \phi + \text{leptons}$ are relevant. The contribution of $\Phi_4$ can be parametrized in the form of the corresponding Wilson coefficient as shown below,
\begin{equation}
\text{Br}(B_{(s)} \to M^{(*)} \ell_1^+ \ell_2^-) = a+ b \, \text{Re}\{C_9\} + c \, |C_9|^2,
\end{equation}
where 
\begin{equation}
C_9 = C_{10} = \frac{\sqrt{2}\pi}{\alpha G_F V_\textsc{ckm}^{tb} (V_\textsc{ckm}^{ts})^*} \frac{c_4^{q_i \ell_1}(c_4^{q_j \ell_2})^*}{4M_{\Phi_4}^2}.
\end{equation}
The predictions on the branching fractions depend on the hadronic from factors which generically suffer from large uncertainties. In this work we have used the parametrization and values for the form factors presented in Refs.~\cite{Bailey:2015dka,Bharucha:2015bzk}. 
In Table~\ref{tab:mesondecays} we list several predictions for different $q^2$ bins. 
Particularly interesting are the decay channels with two electrons in the final state, since the Wilson coefficient entering these processes is fixed by Eq.~\eqref{eq:conditionangles}, i.e. by the neutral anomalies. We list their branching ratios as a function of the SM prediction:
\begin{eqnarray}
\text{Br}(B \to K e^+ e^-) &\simeq  &1.15 \times \text{Br}(B \to K e^+ e^-)_\text{SM},  \qquad  \, \,  \text{ for } q^2 \supset [1.1,6] \text{ GeV}^2\\ 
\text{Br}(B \to K^* e^+ e^-) &\simeq & 1.23 \times \text{Br}(B \to K^* e^+ e^-)_\text{SM},   \qquad  \text{ for } q^2 \supset [0.045,6] \text{ GeV}^2 \\
 \text{Br}(B_s \to \phi \, e^+ e^-)  & \simeq & 1.27 \times  \text{Br}(B_s \to \phi \, e^+ e^-)_\text{SM}, \qquad \,  \text{ for } q^2 \supset [1.1,6] \text{ GeV}^2 . 
\end{eqnarray}
We note that the above processes are also sensitive to long distance effects involving the charm quark that could alter the predictions. Recently, LHCb has measured for first time the semileptonic decay of B mesons to electrons~\cite{LHCb:2021lvy}. Given that the error of the experimental measurements is larger than the $20\%$ of the central value, and the uncertainties in the SM predictions are also large, the above predictions can be consistent with the current bounds.

\end{itemize}

\FloatBarrier

\section{Implications for the quark-lepton unification scale}
%
Using the textures for $V$ and $V_c$ discussed in Sec.~\ref{sec:MQL}, one can study the implications for all interactions of the leptoquarks present in the theory involving these mixing matrices. 
Typically, in theories based on the Pati-Salam gauge symmetry, the vector leptoquark $X_\mu \sim (\mathbf{3},\mathbf{1},2/3)_\text{SM}$ has to be heavy, $M_X > 10^3$ TeV, in order to satisfy the stringent experimental bound from 
$K_L \to e^\pm \mu^\mp$~\cite{Valencia:1994cj,Smirnov:2007hv}. The latter is true if one neglects the mixing between the fermions. However, the lower bound on the vector leptoquark mass depends on the mixing between the quarks and leptons. The relevant interactions for the vector leptoquark can be written as
\begin{eqnarray}
&&\bar{d}_i e_j X_\mu: \hspace{0.5cm} i \frac{g_4}{\sqrt{2}}  \left[ V^{ij} P_R  + (V_c^*)^{ij} P_L  \right] \gamma_\mu.
\label{eq:VLQint}
\end{eqnarray}
The vector leptoquark $X_\mu$ also interacts with the up quarks and the neutrinos, but their mixing matrices are not constrained and, in general, there is freedom to evade the strongest experimental bounds on the processes that it can mediate. However, we have information about the textures of the $V$ and $V_c$ matrices, i.e. the mixing matrices for the charged leptons and down quarks, from the experimental constraints on the Yukawa interactions between the scalar leptoquarks and the fermions. We will therefore focus on their effect on the vector leptoquark through the interactions in Eq.~\eqref{eq:VLQint}.

By looking at the generic expression for the leptonic decay $M \to \bar \ell_i \ell_j$, where $M \equiv \bar d_k d_l$ is a meson made of down-type quarks and $\ell_i \neq \ell_j$ are light leptons ($e$ and $\mu$), 
\begin{eqnarray}
\text{Br}(M \to  \ell_i^- \ell_j^+) = \frac{\tau_{M}}{32\pi}\frac{f_M^2}{m_M^3} && (m_M^2-m_\mu^2)^2 \left( \left | (C_9 - C_9')^{\bar \ell_i \ell_j \bar d_k d_\ell}(m_{\ell_1}-m_{\ell_2})+(C_S-C_S')^{\bar \ell_i \ell_j \bar d_k d_\ell}\frac{m_{M}^2}{m_{d_k}+m_{d_l}}\right|^2 \right. \nonumber \\
&&\left. + \left| (C_{10}-C_{10}')^{\bar \ell_i \ell_j \bar d_k d_\ell}(m_{\ell_1}+m_{\ell_2})+(C_P-C_P')^{\bar \ell_i \ell_j \bar d_k d_\ell}\frac{m_{M}^2}{m_{d_k}+m_{d_l}}\right|^2\right),
\label{eq:KLongGen}
\end{eqnarray}
where the wilson coefficients, defined as
\begin{equation}
\begin{split}
{\cal L}_\text{eff} &\supset [C_{9}^{(')}]^{\bar \ell_i \ell_j \bar d_k d_l} (\bar \ell_i \gamma_\mu \ell_j)(\bar d_k \gamma^\mu P_{L(R)} d_l) + [C_{10}^{(')}]^{\bar \ell_i \ell_j \bar d_k d_l} (\bar \ell_i \gamma_\mu \gamma_5 \ell_j) (\bar d_k \gamma^\mu P_{L(R)} d_l)\\
&+[C_S^{(')}]^{\bar \ell_i \ell_j \bar d_k d_l} (\bar \ell_i \ell_j)(\bar d_k P_{R(L)}d_l) + [C_P^{(')}]^{\bar \ell_i \ell_j \bar d_k d_l} (\bar \ell_i \gamma_5 \ell_j)(\bar d_k P_{R(L)} d_l) + \text{h.c.},
\end{split}
\end{equation}
after integrating out the vector leptoquark $X_\mu$ are given by
\begin{equation}
\begin{split}
&[C_9]^{\bar \ell_i \ell_j \bar d_k d_l} \! = -[C_{10}]^{\bar \ell_i \ell_j \bar d_k d_l} \! = \frac{g_4^2}{4M_X^2}(V^{li})^*V^{kj},  \quad
[C_9']^{\bar \ell_i \ell_j \bar d_k d_l} \! = [C_{10}']^{\bar \ell_i \ell_j \bar d_k d_l} \! = \frac{g_4^2}{4M_X^2}V_c^{li}(V_c^{kj})^*, \\
&[C_P]^{\bar \ell_i \ell_j \bar d_k d_l} \! = - [C_S]^{\bar \ell_i \ell_j \bar d_k d_l}  \! = \frac{g_4^2}{2M_X^2}V_c^{li}V^{kj}, \quad [C_P']^{\bar \ell_i \ell_j \bar d_k d_l} \! = [C_S']^{\bar \ell_i \ell_j \bar d_k d_l} \! =  -\frac{g_4^2}{2M_X^2}(V^{li})^*(V_c^{kj})^*.
\end{split}
\end{equation}
From Eq.~\eqref{eq:KLongGen} we note that the contribution from the scalar currents is largely enhanced (a factor of $\sim 600$ in the process of $K_L \to \mu^\pm e^\mp$ at the level of the amplitude squared) by the ratio of the meson and quark masses. Strikingly, the textures in Eqs.~\eqref{eq:Vctexture} and~\eqref{eq:Vtexture} enable the suppression of such dangerous Wilson coefficients by taking the limit $\theta \to 0$, which as we learned from Sec.~\ref{sec:FlavorSignatures} is required by meson mixing constraints. In that case, only the vector Wilson coefficients
\begin{equation}
{C_9'}^{\bar \mu e \bar s d} = {C_{10}'}^{\bar \mu e \bar s d} = \frac{g_4^2}{4M_X^2}V_c^{12}(V_c^{21})^*, \qquad \text{ and } \qquad {C_9'}^{\bar e \mu \bar d s} = {C_{10}'}^{\bar e \mu \bar d s}  = \frac{g_4^2}{4M_X^2}V_c^{21}(V_c^{12})^*,
\end{equation}
are non-suppressed, and the branching ratio $K_L \to \mu^\pm e^\mp$ is given by
\begin{equation}
\text{Br}_{K_L \to \mu^\pm e^\mp}^X \simeq \frac{\tau_{K_L} \pi}{32} \frac{f_K^2}{m_K^3} (m_K^2-m_\mu^2)^2 m_\mu^2 \left(\frac{\alpha_4}{M_X^2}\right)^2 \cos^2 \theta_c,
\end{equation}
which, requiring it to satisfy the experimental bound $\text{Br}(K_L \to \mu^\pm e^\mp) < 4.7 \times 10^{-12}$, allows to relax the generic $M_X \gtrsim 10^3$ TeV bound down to 
\begin{equation}
M_X \gtrsim 74 \text{ TeV} \left(\frac{\alpha_4}{0.118}\right)^{1/2} \left | \frac{\cos \theta_c }{0.1} \right |^{1/2}.
\end{equation}
The above constraint depends on $\cos \theta_c$, which can be as small as $0.04$ according to Eq.~\eqref{eq:conditionangles} and perturbativity of the Yukawa couplings. Furthermore, a small $\cos \theta_c$ is enforced by the strong $B_s-\bar B_s$ constraint, see Eq.~\eqref{eq:BsBsmix}. However, in the limit $\cos \theta_c \to 0$, other bounds such as $\mu \to e \gamma$ will become relevant. In this case, $M_X \gtrsim 220 \text{ TeV} \sqrt{|\sin \theta_c / 0.7|}$ in the limit where $\sin \theta \to 0$.

\section{Summary}
We have discussed the simplest gauge theory for the unification of quarks and leptons that can describe physics at the TeV scale. We have shown 
that the interactions of the scalar leptoquarks present in the theory can be used to explain the flavor anomalies in agreement with all experimental constraints.
We have discussed the correlation between the flavor violating couplings and the fermion masses, showing that the interactions relevant for the down quarks and charged leptons 
are bounded from above by the quark and lepton masses. The minimal theory of quark-lepton unification predicts a correlation between the ratios ${\cal R}_{K^{(*)}}$ and $\text{Br}(\tau \to e \gamma)$ that will allow to test its potential to address the flavour anomalies in a foreseeable future. Strikingly, another implication of the predicted flavour structure of the fermion mixing matrices is that the simplest gauge theory for quark-lepton unification can be realized at the $100$ TeV scale in agreement with all experimental constraints.

\vspace{1.0cm}
{\small{{\textit{Acknowledgments: We thank A. D. Plascencia for discussions and detailed comments on the draft. C.M. thanks Mark B. Wise for helpful discussions. The work of C.M. is supported by the U.S. Department of Energy, Office of Science, Office of High Energy Physics, under Award Number DE-SC0011632 and by the Walter Burke Institute for Theoretical Physics.}}}}

\appendix

\section{Charged Flavour Anomalies}
\label{app:Charged}

Three different experiments have also been reporting anomalies in the ratios~\cite{Lees:2012xj,Lees:2013uzd,Aaij:2015yra,Huschle:2015rga,Hirose:2016wfn,Hirose:2017dxl,Aaij:2017uff,Aaij:2017deq,Abdesselam:2019dgh},
\begin{equation}
{\cal R}_{D^{(*)}} \equiv \frac{\text{Br}(B \to D^{(*)} \tau \bar{\nu}) }{ \text{Br}(B \to D^{(*)} \ell \bar{\nu})},
\end{equation} 
with $\ell=e,\mu$. The world averages for their experimental measurements given by the Heavy Flavour Averaging Group (HFLAV) are
\begin{equation}
{\cal R}_D=0.339\pm 0.026 \pm 0.014, \qquad \textrm{and} \qquad {\cal R}_{D^{*}}=0.295 \pm 0.010 \pm 0.010,
\label{eq:ratiosD}
\end{equation}
with correlation -0.38, while the predicted values in the SM are: ${\cal R}_D^{SM}=0.298 \pm 0.003$ and ${\cal R}_{D^*}^{SM}=0.252 \pm 0.005$~\cite{HFLAV:2019otj}. The reported combined tension for the ${\cal R}_{D^{(*)}}$ measurements with respect to the SM predictions is about $3.4~\sigma$~\cite{HFLAV:2019otj}.

The above processes involve $b \to c$ transitions and in the literature are commonly referred as charged anomalies. As we show in the following, the minimal theory for quark-lepton unification can also provide a solution for them.
The $\phi_4^{2/3}$ effective interactions contributing to $b \to c$ transitions are listed below,
\begin{equation}
{\cal H}_\text{eff}^{b \to c} \supset 
 \frac{c^{il}_2 (c_4^*)^{jk}}{2M_{\Phi_4}^2} \left[ (\bar u_R^i d_L^j)(\bar e_R^k \nu_L^l) + \frac{1}{4}(\bar u_R^i \sigma^{\mu \nu} d_L^j) (\bar e_R^k \sigma_{\mu \nu} \nu_L^l) \right] + \text{h.c.},
\end{equation}
where $c_2=U_C^T Y_2^T N$. Notice that the $Y_2$ Yukawa matrix is required, whose entries are not restricted as $Y_4$ because the Dirac masses of the neutrinos are unknown. Reading from the above equation the Wilson coefficients in the basis 
\begin{equation}
\begin{split}
{\cal H}_\text{eff}^{b \to c} \supset  \frac{4 G_F}{\sqrt{2}} V_\textsc{ckm}^{cb}  &\left[ (\bar c \gamma^\mu P_L b)(\bar \tau \gamma_\mu P_L \nu) + C^S_{LL} (\bar c P_L b)(\bar \tau P_L \nu) + C^T_{LL} (\bar c \sigma^{\mu \nu} P_L b)(\bar \tau \sigma_{\mu \nu} P_L \nu) \right] 
+ \text{h.c.},
\end{split}
\end{equation}
we can identify two independent degrees of freedom: the real and the imaginary part of the following Wilson coefficient,
\begin{equation}
C^S_{LL} = 4 \, r \, C^T_{LL} = \! \frac{\sqrt{2}}{8G_F V_\textsc{ckm}^{cb}} \frac{c_2^{23}(c_4^{33})^*}{M_{\Phi_4}^2},
\end{equation}
where $r\sim 2$ takes into account the running of this Wilson coefficient between $1$ TeV and the bottom mass scale~\cite{Freytsis:2015qca,Dorsner:2016wpm}.

\begin{figure}[t]
\includegraphics[width=0.49\linewidth]{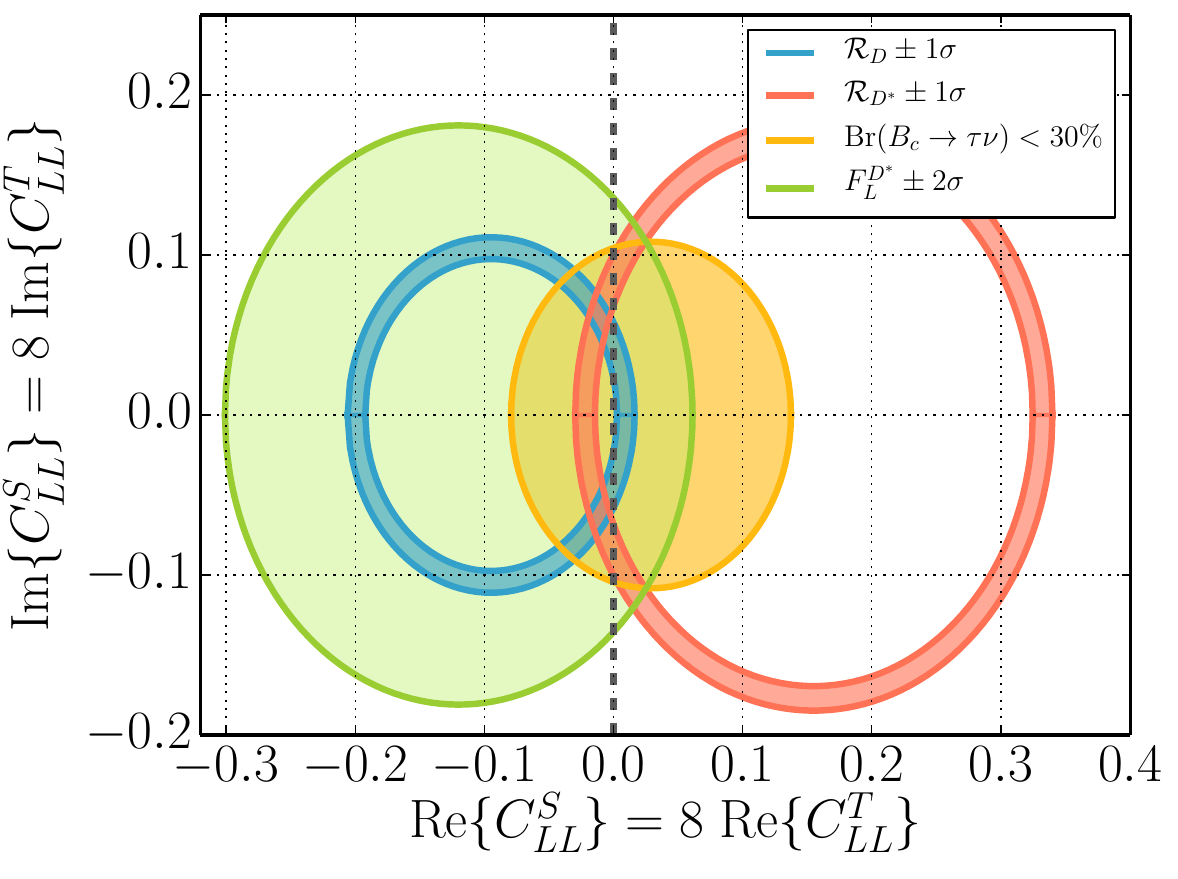}
\includegraphics[width=0.46\linewidth]{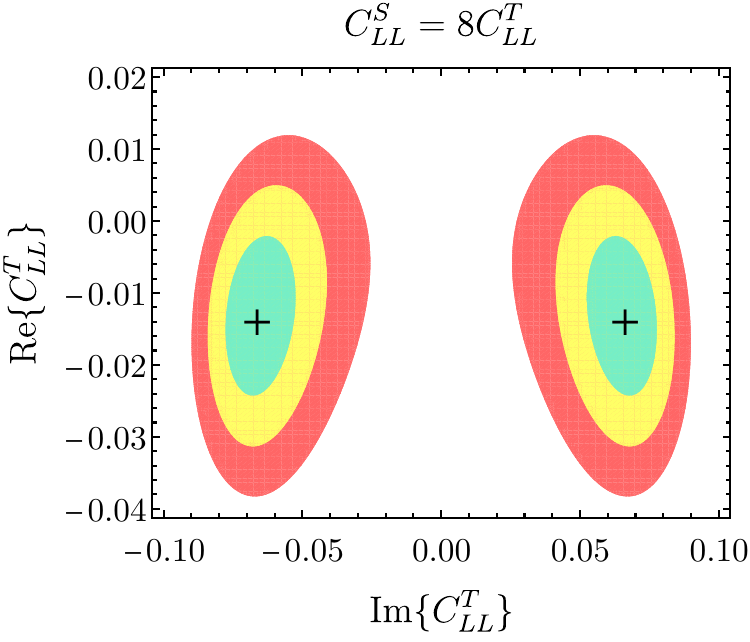}
\caption{Left panel: Parameter space satisfying different charged anomalies in the $\text{Re}\{C_{S}^{LL}\} =8 \, \text{Re}\{C_{T}^{LL}\}$ and $\text{Im}\{C_{S}^{LL}\}=8 \, \text{Im}\{C_{T}^{LL}\}$ plane. In blue, the experimental result for ${\cal R}_D^\text{exp}$ at $1\, \sigma$, in red ${\cal R}_{D^*}^\text{exp}$ at $1\, \sigma$, in red the $\text{Br} (B_c \to \tau \nu) < 30\%$ constraint, and in green the $F_L^{D^*}$ at $2\, \sigma$. Right-panel: Allowed regions at $68\%$, $95\%$ and $99.7\%$ C.L. (green, yellow and red, respectively) in the $\text{Im}\{C_{S}^{LL}\} =8 \, \text{Im}\{C_{T}^{LL}\}$ and $\text{Re}\{C_{S}^{LL}\}=8 \, \text{Re}\{C_{T}^{LL}\}$ plane, for the fit including the ratios ${\cal R}_{D^{(*)}}$ and $F_L^{D^*}$.}
\label{fig:charged}
\end{figure}

The left panel of Fig.~\ref{fig:charged} shows that the ratios ${\cal R}_{D}$ and ${\cal R}_{D^*}$ can be satisfied at the $1 \, \sigma$. However, when considering the longitudinal $D^*$ polarization, $F_L^{D^*}$, the experimental measurement by Belle~\cite{Belle:2019ewo} can only be accommodated at $\sim 2\, \sigma$, as noted in Ref.~\cite{Mandal:2020htr}. By fitting ${\cal R}_{D}$, ${\cal R}_{D^*}$ and $F_L^{D^*}$ to their experimental measurements (the ratios are given in Eq.~\eqref{eq:ratiosD} and $F_L^{D^*} = 0.60  \pm 0.08 \pm 0.04$~\cite{Belle:2019ewo}) allowing the real and imaginary parts of $C^S_{LL} = 8 C^T_{LL}$ to vary, we obtain a $\chi^2 \simeq 2.3$ (per one degree of freedom). The latter reflects the tension in accommodating the experimental measurement $F_L^{D^*}$, which cannot be addressed together with the ratios under any kind of new physics under well-motivated assumptions~\cite{Murgui:2019czp}. In the fit we used the form factors as treated in Ref.~\cite{Murgui:2019czp}, where the formulae for the relevant $b \to c$ processes can be found, and assumed the indirect bound $\text{Br}(B_c \to \tau \nu) \lesssim 30\%$~\cite{Beneke:1996xe, Celis:2016azn,Alonso:2016oyd}, although the minimum does not saturate it as the left panel of Fig.~\ref{fig:charged} shows. The preferred solutions of the fit are,\footnote{We fit the new physics Wilson coefficients to the experimental ratios ${\cal R}_{D^{(*)}}$ in Eq.~\eqref{eq:ratiosD}, which have been obtained by assuming the SM nature of the semitauonic decays. As pointed out in Ref.~\cite{Bernlochner:2020tfi}, this leads to biases which in the case of the leptoquark of interest, $\Phi_4$, may be non-negligible. Avoiding the possible biases in fit would require properly reweighting the ratios in the context of the new physics scenario, which is out of the scope of this work. We rather aim to show that $\Phi_4$ has the potential of addressing such deviations in the minimal theory of quark-lepton unification.}
\begin{equation}
\text{Re}\{ C_T^{LL} \} =  \frac{\text{Re}\{C_{S}^{LL}\}}{8} = -0.013 \pm 0.007, \quad \text{ and } \quad \text{Im}\{ C_T^{LL}\} = \frac{\text{Im}\{C_{S}^{LL}\}}{8} = 0.066^{+0.008}_{-0.009}.
\end{equation}
Therefore, apart from a complex phase as was noted in Ref.~\cite{Sakaki:2013bfa}, a non-zero $c_2^{23}$ matrix entry is needed in order to address the charged anomalies in this theory.

\bibliography{QLUnification.bib}

\end{document}